\documentclass[journal]{IEEEtran}

\IEEEoverridecommandlockouts
\usepackage{amsmath,amsfonts}
\usepackage{bbding}
\usepackage{algorithmic}
\usepackage{algorithm}
\usepackage{array}
\usepackage[caption=false,font=normalsize,labelfont=sf,textfont=sf]{subfig}
\usepackage{textcomp}
\usepackage{stfloats}
\usepackage{url}
\usepackage{verbatim}
\usepackage{graphicx}
\usepackage{CJKutf8}
\usepackage{balance}
\usepackage{wasysym}
\usepackage{siunitx}
\usepackage{booktabs}
\usepackage{color, xcolor}
\usepackage{tikz,xcolor}
\usepackage{lineno} 
\usepackage{soul}
\usepackage{multirow}
\usepackage{caption}
\usepackage{cuted}
\usepackage[numbers,sort&compress]{natbib} 
\usepackage{hyperref}

\definecolor{lime}{HTML}{A6CE39}
\DeclareRobustCommand{\orcidicon}{
\begin{tikzpicture}
\draw[lime, fill=lime] (0,0)
circle[radius=0.16]
node[white]{{\fontfamily{qag}\selectfont \tiny \.{I}D}};
\end{tikzpicture}
\hspace{-2mm}
}
\foreach \x in {A, ..., Z}{%
\expandafter\xdef\csname orcid\x\endcsname{\noexpand\href{https://orcid.org/\csname orcidauthor\x\endcsname}{\noexpand\orcidicon}}
}

\graphicspath{{./figure/}}    

\begin{document}


\switchlinenumbers	
\bstctlcite{IEEEexample:BSTcontrol}
\title{NeuroPDE$^+$: A Scalable Neuromorphic PDE Accelerator Based on Spintronic and Ferroelectric Devices}

\author{Siqing Fu$^1$\hspace{-1.5mm}\orcidA{}, Lizhou Wu$^1$\hspace{-1.5mm}\orcidC{}, Tiejun Li$^{\ast}$\hspace{-1.5mm}\orcidH{}, Xuchao Xie$^{\ast}$\hspace{-1.5mm}\orcidJ{},\\  Sheng Ma\hspace{-1.5mm}\orcidE, Jianmin Zhang\hspace{-1.5mm}\orcidG{}, Wei Chen \hspace{-1.5mm}\orcidL{}, Yunping Zhao \hspace{-1.5mm}\orcidK{}

    \thanks{$^1$These authors contributed equally to this work.}
    \thanks{Manuscript received 5 November 2025; revised 5 June 2026 and 18 August 2026; accepted 12 September 2026. This work is supported in part by the NSFC (62472435, 62172430, 62304257), the STIP of Hunan Province 2022RC3065, the Foundation of PDL 2023-JKWPDL-02, and the Foundation of NUDT (25-ZZCX-JDZ-16).
    ($^{\ast}$Corresponding author: Tiejun Li and Xuchao Xie.)}

    \thanks{Siqing Fu, Lizhou Wu, Tiejun Li, Xuchao Xie, Chunyuan Zhang, Sheng Ma, Jianmin Zhang are with the College of Computer Science and Technology, National University of Defense Technology, Changsha 410073, China. Siqing Fu is also with the Jiangnan Institute of Computing Technology, Wuxi 214026, China. (e-mail: \{fusiqingnudt, lizhou.wu, tjli, xiexuchao, masheng, jmzhang, chenwei, zhaoyunping\}@nudt.edu.cn).}


}




\maketitle

\begin{abstract}

The pursuit of high-performance PDE solvers rests on three fundamental challenges: (i) the curse of dimensionality in kinetic and financial equations, (ii) the poor extrapolation of purely data-driven surrogates, and (iii) the widening gap between algorithm design and hardware specialization. To overcome these challenges, we present NeuroPDE$^+$, a scalable neuromorphic PDE solver design based on spintronic and ferroelectric devices for accelerating PDE solutions.  NeuroPDE$^+$ consists of two dedicated units:  a diffusion tracking unit (DTU), which emulates random walks on Markov chains through activations between hardware neurons, and a scattering tracking unit (STU), which samples non-local jumps via a multi-level probability tree. 
System-level simulations suggest that NeuroPDE$^+$ achieves a squared error below 1e-2 in steady-state heat equation and particle transport problems. Simulation results further indicate that the DTU  achieves up to a 315$\times$ performance gain over previous neuromorphic processors, and that the STU achieves a 1000$\times$ speedup compared to a general-purpose CPU. Co-designing algorithm and hardware with intrinsic stochasticity and non-volatile in-memory computing, NeuroPDE$^+$ preliminarily explores a new paradigm for efficient and scalable neuromorphic PDE solvers. This approach could pave the way for probabilistic computing architectures in large-scale scientific simulations.

\end{abstract}

\begin{IEEEkeywords}
Neuromorphic, PDE Solver, Monte Carlo, MTJ, FTJ.
\end{IEEEkeywords}
\vspace{-20pt} 
\section{Introduction}
\label{Intro}

\IEEEPARstart{P}{\textit{artial}} \textit{differential equations} (PDEs) describe relationships between variables, their partial derivatives, and unknown functions, with broad applications in scientific and engineering fields such as materials science, aerospace engineering, and fluid mechanics \cite{hu2020heat, gao2021bi, Hanindhito2021, Gourounas2025}. Solving PDEs remains a fundamental challenge in scientific computing due to their inherent complexity, driving continuous research efforts. Traditional solver methods include analytical approaches \cite{henner2019partial}, finite difference methods, finite element techniques \cite{Claes1990}, and spectral methods \cite{townsend2015automatic}, all of which typically require domain discretization followed by numerical solutions at grid points. While most PDEs of practical interest are two- or three-dimensional, problems involving parametric or stochastic PDEs can become high-dimensional, where traditional numerical methods suffer from the curse of dimensionality and become computationally prohibitive \cite{hu2024tackling}.

Recent advances in data-driven artificial intelligence, particularly in deep learning, have motivated the development of novel methods for solving PDEs \cite{blechschmidt2021three,karniadakis2021physics,meng2023pinn, Pestourie2023}. By incorporating physical constraints into neural network loss functions, trained models can, in principle, learn to approximate PDE solutions, yet their effectiveness varies significantly with problem complexity. However, neural PDE solvers relying on equation-constrained training or pre-generated data often exhibit limited extrapolation power, as training data rarely spans the full range of practical inputs and parameter regimes \cite{Jiang2023}. To bypass this and the associated training cost, researchers have begun exploring unsupervised probabilistic neural solvers.

The novel \textit{Monte Carlo} (MC) random walk solver \cite{severa2018spiking, smith2020solving, smith2022neuromorphic,zhang2023monte} employs probabilistic PDE representations to train unsupervised models for solving convection-diffusion, Allen-Cahn, and Navier-Stokes equations. However, its efficiency is limited by the von Neumann architecture, where random particle tracking causes irregular branching and memory access \cite{ma2021}. Neuromorphic computing, which mimics the architecture of the brain, has been explored as a potential solution to this bottleneck. For instance, implementations on Loihi \cite{davies2018loihi} and TrueNorth \cite{Akopyan} have demonstrated significantly reduced energy consumption. However, a key limitation of these CMOS-based neuromorphic architectures is their lack of inherent randomness, which is crucial for efficient MC methods. Consequently, despite their energy efficiency, they have not yet demonstrated significant performance advantages over conventional processors for this specific application \cite{smith2022neuromorphic}. The design of novel PDE solvers is challenged by the need to overcome the curse of dimensionality inherent in traditional methods while simultaneously addressing the prohibitive training costs of neural network solvers and the architectural limitations of MC solvers.

To address the limitations of existing approaches, we propose NeuroPDE$^+$, a scalable MC neural PDE accelerator that uses spintronic stochasticity and ferroelectric memristors. Here ``scalability" refers to the ability of the architecture to handle broader PDE classes, including both local diffusion and nonlocal scattering, rather than merely increasing grid size or dimensions. The design employs \textit{magnetic tunnel junctions} (MTJs) for physical randomness and \textit{ferroelectric tunnel junctions} (FTJs) for in-memory computing. The architecture that combines \textit{diffusion tracking unit} (DTU) and \textit{scattering tracking unit} (STU) provides a tight integration between probabilistic algorithms and their hardware implementation, where the DTU is dedicated to tracking diffusion processes via random walk models, while the STU extends the capability to more complex stochastic models that include scattering events. Our SPICE and system simulation results suggest that NeuroPDE$^+$ can solve PDEs in an energy-efficient manner, achieving a squared error below 1e-2. Simulation results also suggest that, compared to CMOS neuromorphic solvers on Loihi and TrueNorth chips, the DTU may achieve up to a 315$\times$ speedup and up to a 29.8$\times$ energy-efficiency improvement for heat equations, and that the STU may show up to three orders of magnitude performance gain and up to six orders of magnitude energy advantage over conventional processors in particle transport solutions, which are upper-bound estimates.
The main contributions of this paper are as follows:

\begin{enumerate} 
    \item We propose a DTU design with hardware-implemented neurons and synapses to accelerate stochastic particle diffusion tracking, which may help offload workloads and alleviate bottlenecks in general-purpose processors.
    \item We propose an STU that generates random sequences to track sampling distributions, aiming to enable the simulation and tracking of complex stochastic processes.
    \item We develop a custom system simulator to evaluate the PDE-solving capability and efficiency of NeuroPDE$^+$; simulation results suggest the potential of spintronic devices for next-generation stochastic computing architectures.
\end{enumerate}

The remainder of this paper is organized as follows. Section~\ref{Back} provides a background on the fundamental principles of MC methods and introduces the novel spintronic and ferroelectric devices used in our circuit design. Section~\ref{Design} details the architecture and design of NeuroPDE$^+$, providing a comprehensive description of both the DTU and the STU. Section~\ref{Experiments} details the experiments, including circuit-level and system-level simulations. Section~\ref{Related} compares the DTU and the STU with their respective related works. Section~\ref{Discussion} addresses the limitations and suggests directions for future research. Finally, Section~\ref{Conclusion} concludes this paper.

\vspace{-10pt} 
\section{Background}
\label{Back}

\subsection{Monte Carlo Method for Solving Diffusion Processes}
\label{Back:MC}

The MC method \cite{MASUDA20171} relates particle trajectory simulations to PDE solutions by capturing stochastic behaviors, such as Brownian motion associated with thermal diffusion. This method involves spatial discretization, the construction of Markov chains to simulate particle movements, and MC sampling to approximate PDE solution spaces.

The Feynman-Kac formula \cite{Hu2019} establishes a fundamental correspondence between a class of PDEs and stochastic processes, providing a probabilistic representation for the solutions of deterministic PDEs. This principle offers a powerful bridge from the world of differential operators to the world of random paths, with profound implications for both theoretical analysis and numerical computation.

The PDEs amenable to this approach can be expressed in the following general form:
\begin{equation}
\frac{\partial u}{\partial t}+\mu  (x,t)\frac{\partial u}{\partial x}+\frac{1}{2}\sigma ^{2}(x,t) \frac{\partial^2u}{\partial x^2}+f(x,t)u=0, 
\label{equPDE}
\end{equation}
where $u=u(x,t)$ is the function to be solved, $\mu(x,t)$ and $\sigma(x,t)$ represent the drift and diffusion terms, and $f(x,t)$ is a given function.

The Feynman-Kac formula states that the solution $u=u(x,t)$ to this PDE is given by the conditional expectation of a functional derived from an associated stochastic process:
\begin{equation}
u(x,t)=\mathbb{E} [\phi(X_{T} )\mid  X_{t} =x] ,
\label{equresolve}
\end{equation}
where $X_{T}$ denotes the stochastic diffusion process.

For computational purposes, the continuous process must be discretized. We model particle motion using a Markov chain on a uniform spatial grid with spacing $\Delta x$. Over a time step $\Delta t$, a particle at node $i$ jumps left or right with equal probability $P_g$, or stays with probability $P_s$, satisfying:
\begin{equation}
P_s + 2P_g = 1.
\label{eq:norm}
\end{equation}

The continuous SDE $dX_t = \mu\,dt + \sigma\,dW_t$ has local moments:
\begin{equation}
\mathbb{E}[\Delta X] = \mu\Delta t, \qquad 
\mathbb{E}[(\Delta X)^2] = \sigma^2\Delta t + (\mu\Delta t)^2.
\label{eq:cont_moments}
\end{equation}

The discrete displacement $\Delta X_d \in \{-\Delta x,\,0,\,+\Delta x\}$ has moments:
\begin{equation}
\mathbb{E}[\Delta X_d] = 0, \qquad 
\mathbb{E}[(\Delta X_d)^2] = 2P_g(\Delta x)^2.
\label{eq:disc_moments}
\end{equation}

Matching the variance (the drift $\mu$ is absorbed into the Feynman-Kac weight) and using Equation (\ref{eq:norm})
\begin{equation}
P_g = \frac{\sigma^2\Delta t}{2\,\Delta x^2}, \qquad
P_s = 1 - \frac{\sigma^2\Delta t}{\Delta x^2},
\label{eq:transition}
\end{equation}
with the stability condition $\sigma^2\Delta t \leq \Delta x^2$. Equation (\ref{equresolve}) is then evaluated by averaging over simulated random walks weighted by $\exp(\int f\,ds)$.


\begin{figure}[t]
\vspace{-10pt}
\centering
\includegraphics[width=0.74\linewidth]{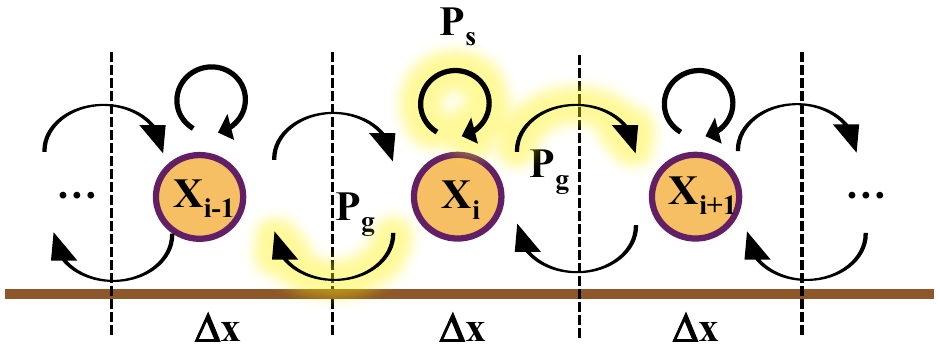}
\caption{Constructed Markov chain model for diffusion processes.} 
\label{fig1}
\vspace{-20pt} 
\end{figure}

Previous research has established that this random walk-based MC PDE solver can solve various equations including the convection-diffusion, Allen-Cahn, and Navier-Stokes equations in an unsupervised manner \cite{zhang2023monte}. However, the Feynman-Kac formula is limited to diffusion processes with local jumps, which model continuous-path processes like Brownian motion.

\subsection{Monte Carlo Method for Solving Scattering Processes}

Many physical systems (e.g. neutron transport, radiation transfer) involve nonlocal jumps requiring integro-differential equations like the Boltzmann equation \cite{Cercignani1988}. For example, particle transport involves scattering mechanisms with nonlocal state transitions, requiring sampling from probability distributions:
\vspace{-8pt}
\begin{equation}
\mathcal{S}\psi(\mathbf{r}, \Omega) = \int_{4\pi} \Sigma_s(\mathbf{r}, \Omega' \rightarrow \Omega) \psi(\mathbf{r}, \Omega') \, d\Omega',
\label{eq:boltzmann_integral}
\end{equation}
where $\Omega$ is the current direction, $\Omega'$ is the scattered direction, and $\Sigma_s$ is the differential scattering cross-section.

\begin{figure}[t]
\vspace{-10pt}
\centering
\includegraphics[width=0.9\linewidth]{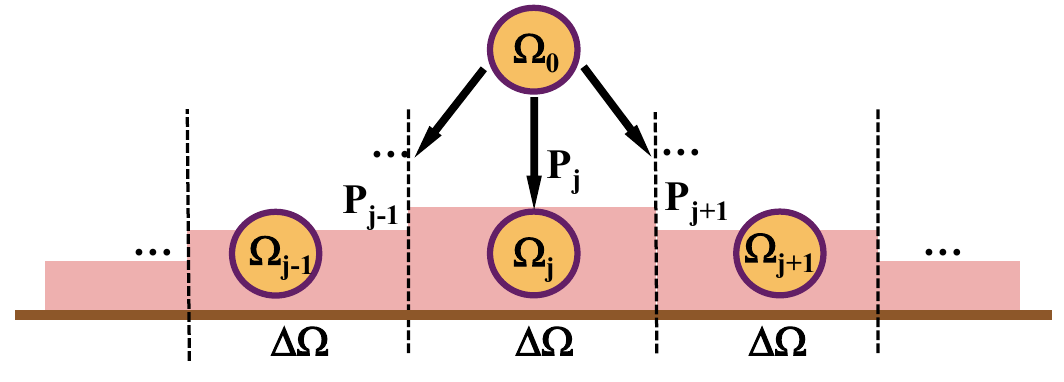}
\caption{Constructed stochastic event sampling model for Scattering processes.} 
\label{figsample}
\vspace{-10pt} 
\end{figure}

Fig.~\ref{figsample} illustrates the stochastic event sampling model for such processes. For a variable $\Omega$ currently in state $\Omega_0$, the potential state space for the next time step is $S_2=\left \{{\Omega_{1},\Omega_{2},...,\Omega_{j},...}  \right \} $. The transition probability from $\Omega_0$ to a state $\Omega_j$ is given by $P_{j}$, which is visually represented by the height of the corresponding discrete-state bar (shown in pink). Consequently, tracking the evolution of $\Omega$ essentially involves sampling from its probability density function to determine subsequent states. In particular, the diffusion process depicted in Fig.~\ref{fig1} represents a special case where sampling is limited to three adjacent states of $\mathrm{X}_{i}$.

To implement hardware-based sampling of scattering directions, the discretized transition probabilities must be mapped onto a circuit structure capable of sampling from arbitrary distributions. We address this challenge in Section~\ref{STUD} by constructing an STU with a conditional probability tree structure.

\vspace{-12pt} 
\subsection{Spintronic and Ferroelectric Devices}

\subsubsection{MTJ Device}
\label{Back:MTJ}
\begin{figure}[t] 
\centering
\includegraphics[width=0.9\linewidth]{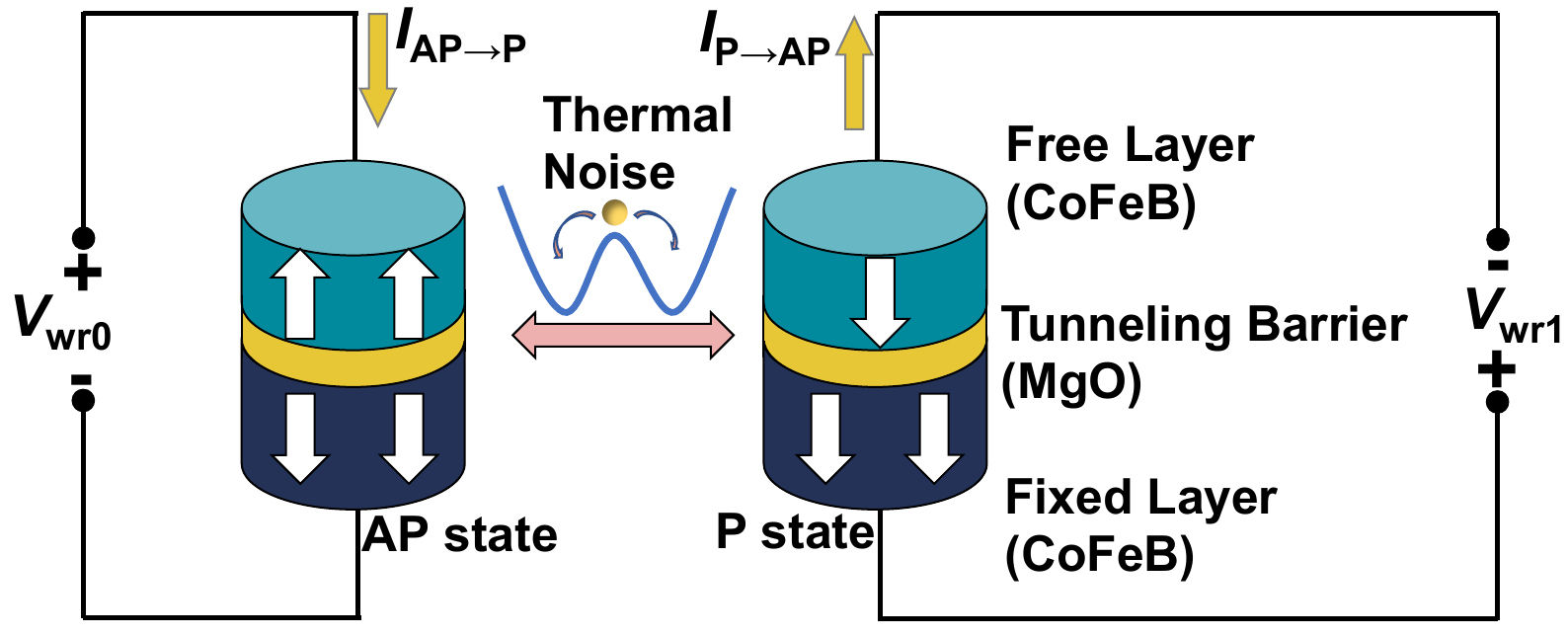}
\caption{MTJ structure and stochastic switching.} 
\label{figmtj}
\vspace{-10pt} 
\end{figure}

Advancements in materials science have established MTJs as promising candidates for MRAM, neuromorphic computing, and probabilistic computing applications, due to their nonvolatility, low power consumption, thermal randomness, and CMOS compatibility. As shown in Fig.~\ref{figmtj}, an MTJ comprises fixed and free magnetic layers separated by a tunneling barrier. The device exhibits two states: high-resistance \textit{anti-parallel} (AP) state (logic ``1") when the magnetizations are anti-parallel, and low-resistance \textit{parallel} (P) state (logic ``0") when aligned. 

Magnetization switching in MTJs involves either \textit{spin-transfer torque} (STT) or \textit{spin-orbit torque} (SOT), with Fig.~\ref{figmtj} illustrating the STT switching process. For an MTJ in the AP state, a voltage $V_{\mathrm{wr0}}$ drives a current $I_{\mathrm{AP}\rightarrow\mathrm{P}}$, which attempts to switch the MTJ to the P state. The reverse current controls the reverse process. Due to thermal noise, the system probabilistically stabilizes in one of the two states, depending on the current amplitude and duration. The switching probability of the MTJ under a bias current of amplitude $I_{\mathrm{wr}}$ and duration $t_{\mathrm{pw}}$ is given by the following equation:
\vspace{-10pt}
\begin{eqnarray}
P(I_{\mathrm{wr}},t_{\mathrm{pw}}) & = & 1-\exp\left(-\frac{t_{\mathrm{pw}}}{\tau}\right), \\
\tau(I_{\mathrm{wr}}) & = & \tau_{0}\exp\left[\Delta\left(1-\frac{I_{\mathrm{wr}}}{I_{c0}}\right)^{2}\right],
\label{equMTJ}
\vspace{-10pt}
\end{eqnarray}
where $\tau$ is the average switching time, $\tau_{0}$ is the attempt time factor, $\Delta$ is the thermal stability factor and $I_{c0}$ is the critical switching current at \SI{0}{K}. By adjusting the write voltage, the pulse amplitude can be controlled to achieve the desired switching probability. The physical randomness of thermal processes in MTJs makes it a promising candidate for an entropy source in MC hardware solvers.

\subsubsection{FTJ Device}
\label{Back:FTJ}
\begin{figure}[t]
\vspace{-10pt} 
\centering
\includegraphics[width=0.9\linewidth]{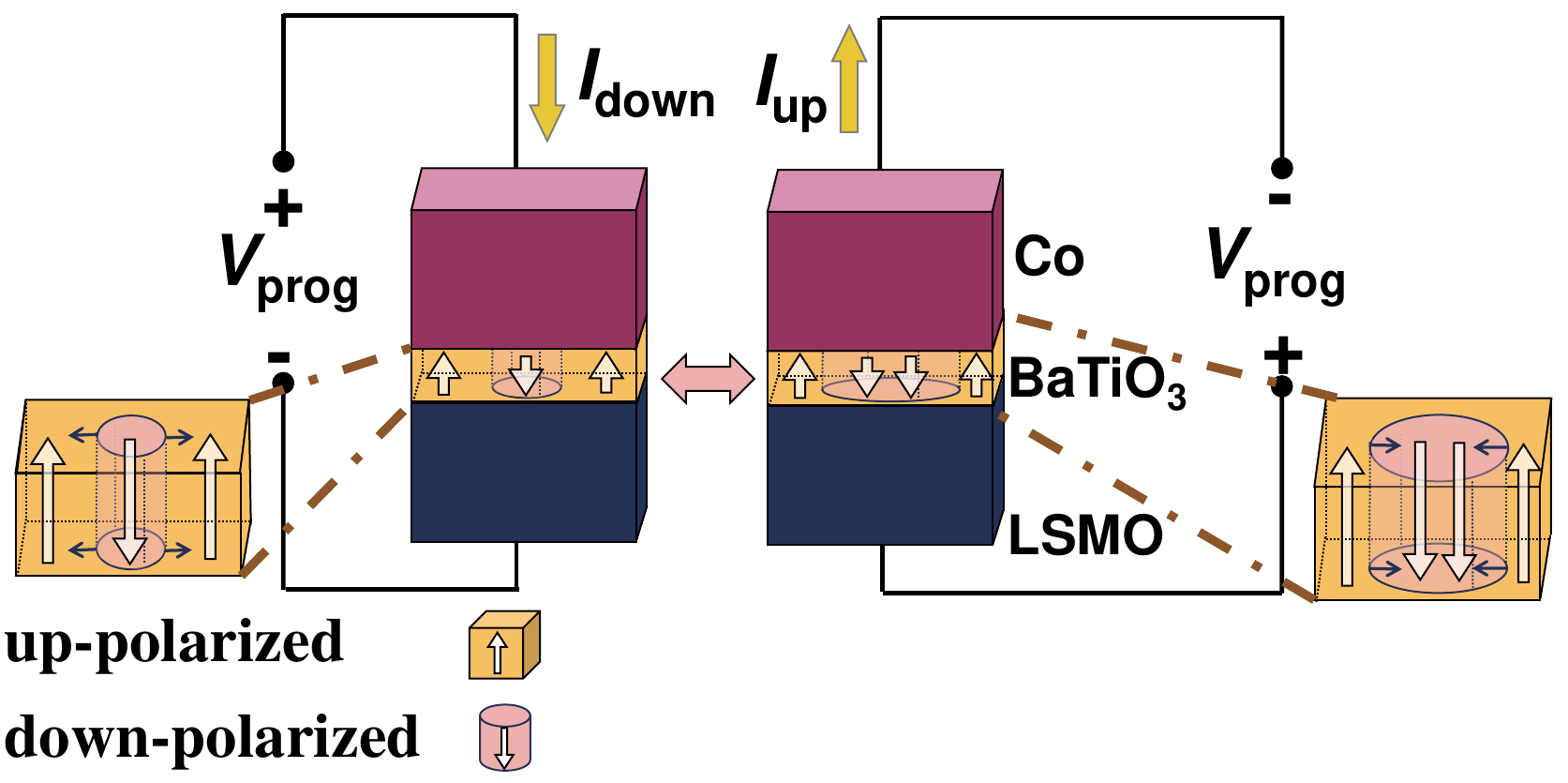}
\caption{FTJ structure and polarization switching process.} 
\label{figftj}
\vspace{-12pt} 
\end{figure}
Circuit randomness is provided by spintronic devices, while FTJs enable probabilistic configurability. Previous works \cite{Fang2024} have observed memristive behavior in FTJs, where resistance changes are induced by ferroelectric polarization switching. Fig.~\ref{figftj} shows an FTJ, which has a structure consisting of Co electrodes, a BTO ferroelectric ultrathin film, and LSMO electrodes, along with the associated polarization switching process. A positive $V_{\mathrm{prog}}$ triggers $I_{\mathrm{down}}$, initiating the nucleation and propagation of domain walls, causing the BTO barrier to flip its polarization and altering the tunneling probability of electrons. The FTJ resistance is controlled by the parallel resistance of oppositely polarized domains, so domain wall motion results in continuous macroscopic resistance changes. Unlike MTJs, which exhibit only two stable resistance states, FTJs offer a continuous range of intermediate resistance values due to the progressive nature of polarization switching and domain wall dynamics. A reverse voltage induces the nucleation and propagation of upward domain walls. The configurable resistance of the FTJ allows precise control of the MTJ switching probability through write voltage adjustment.

\vspace{-10pt} 
\section{Proposed NeuroPDE$^+$ Design}
\label{Design}

\begin{figure}[t]
	\centering
	\vspace{-15pt} 
	\includegraphics[width=7cm]{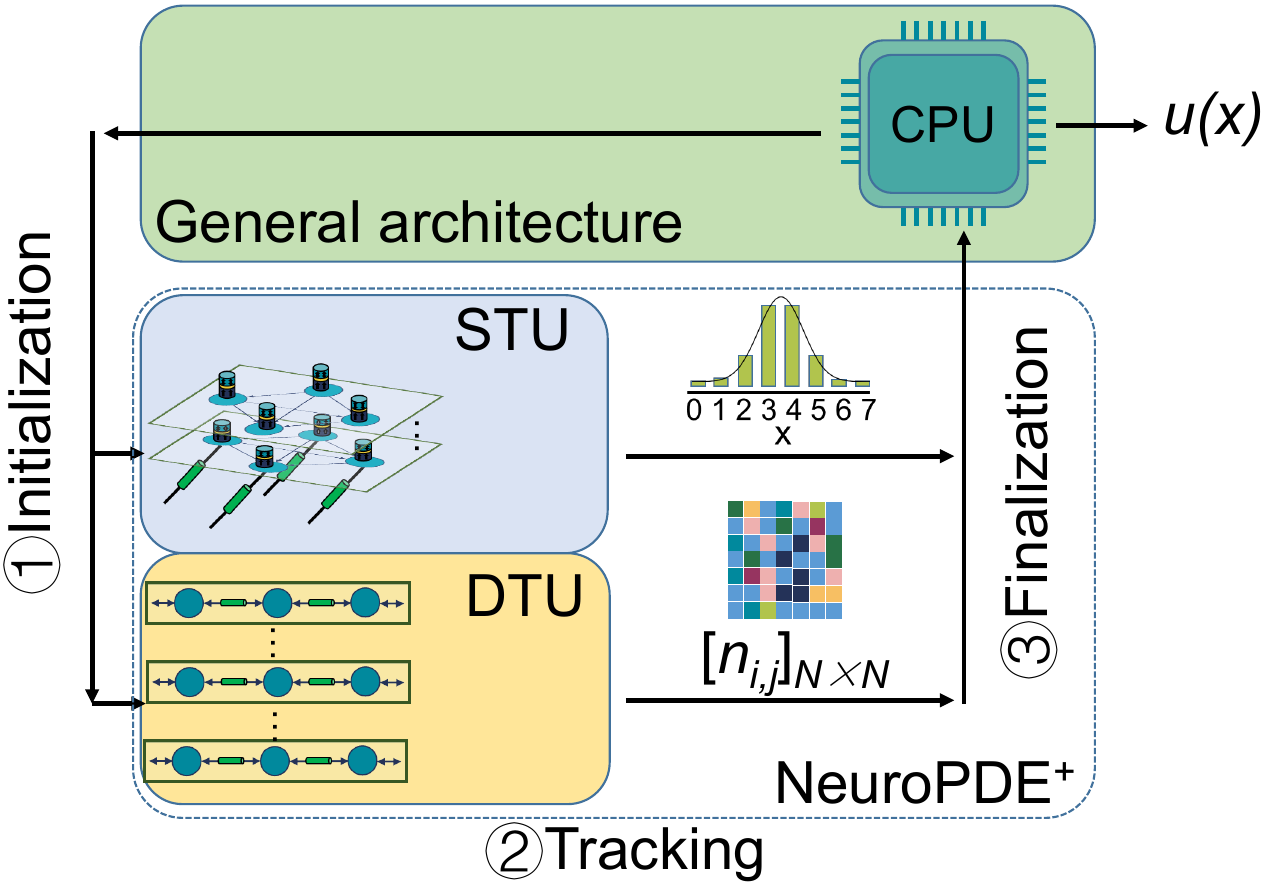}
	\caption{System-Level Architecture of NeuroPDE$^+$: A Monte Carlo PDE Solver.}
	\label{figsys}
	\vspace{-5pt} 
\end{figure}

\subsection{Idea and Goal}

To address the limitations of prior PDE solvers, we propose NeuroPDE$^+$, an extended neuromorphic accelerator for generalized PDE solving. The solution implements two key processing units: a \textit{diffusion tracking unit} (DTU) for diffusion processes modeled by random walks, and a \textit{scattering tracking unit} (STU) for non-local jump processes using probability sampling. As shown in Fig.~\ref{fig1} and Fig.~\ref{figsample}, the Markov chain model and the probabilistic sampling model are mapped to these respective hardware units. A system-level view of the overall architecture and the mapping of MC operations to hardware is provided in Fig.~\ref{figsys}, where the host CPU handles initialization and weight programming, the DTU and STU execute the tracking phase independently on-chip as decoupled units, and results are returned to the CPU for final computation, so they are evaluated separately in this work. The circuit implementation leverages spintronic and ferroelectric device properties to achieve complete MC simulation for PDE solving.

Our main objectives include:
\begin{enumerate} 
    \item \textbf{DTU implementation:} The first objective is to implement the DTU. This is achieved by utilizing the random switching behavior of MTJ devices to simulate discrete random processes and employing ferroelectric devices to control and store transition probabilities. This enables precise neural activation that accurately matches random walk processes.
    \item \textbf{STU implementation:} The second objective involves realizing accurate sampling of complex probability distributions through a multi-level conditional probability tree structure, where MTJ devices provide entropy sources and FTJ programmable resistors dynamically adjust sampling probabilities. This approach enables hardware acceleration of non-local jump processes like particle scattering.
\end{enumerate}

\vspace{-10pt} 
\subsection{Diffusion Tracking Unit Design}

\subsubsection{Design Philosophy}

\begin{figure}[t]
\centering
\vspace{-15pt} 
\includegraphics[width=9cm]{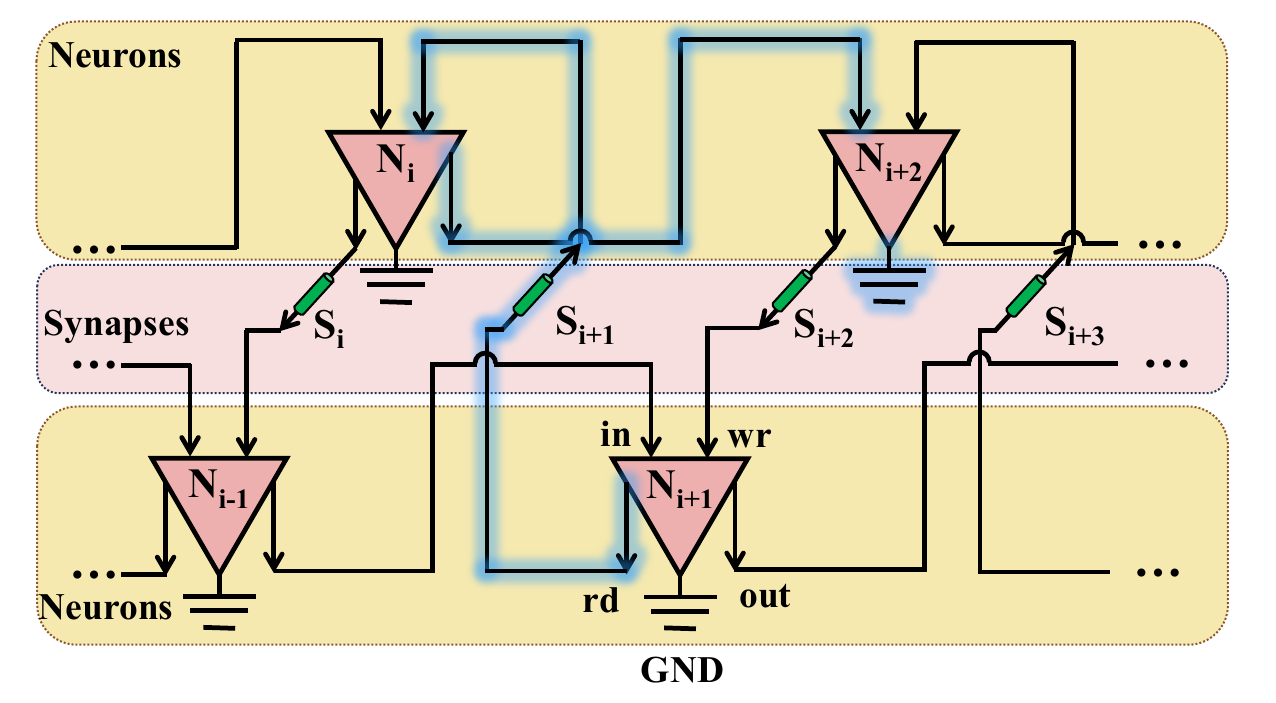}
\caption{DTU design overview: neurons arranged by parity and interconnected via synapses, highlighting activation pathways from neuron N$_{i+1}$ to neighboring neurons.} 
\label{fig4}
\vspace{-5pt} 
\end{figure}

To enable hardware-based tracking of random walks for supporting MC random walk solutions to PDEs, we designed a DTU corresponding to the Markov chain model. Fig.~\ref{fig4} shows the structural design of the DTU. Each neuronal unit $\mathrm{N}_i$ maps to a discrete spatial point $\mathrm{X}_{i}$ in Fig.~\ref{fig1}. The transition probabilities, such as $P_{s}$, are encoded in the distributed synaptic weights across the network. Neurons follow a logical linear sequence but are physically placed in alternating odd and even rows to ensure balanced bidirectional activation. Every neuron has five I/O pins: $\mathrm{in}$, $\mathrm{wr}$, $\mathrm{rd}$, $\mathrm{out}$, and $\mathrm{GND}$.

\begin{figure}[t]
\centering
\includegraphics[width=7.5cm]{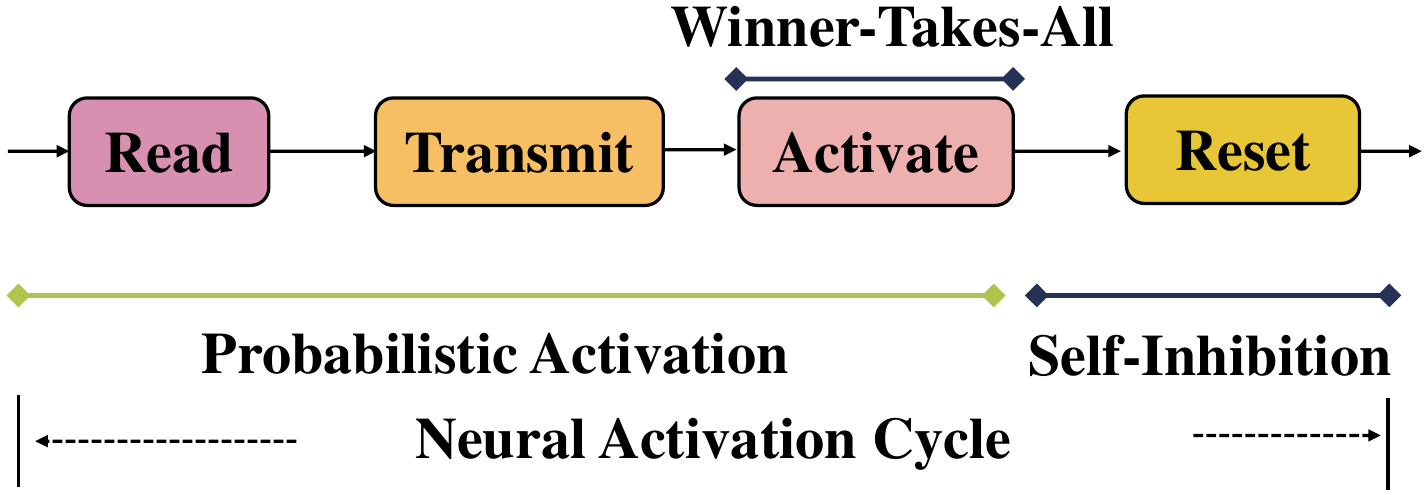}
\caption{A process of neural activation cycle.} 
\label{fig4.5}
\vspace{-15pt} 
\end{figure}

Fig.~\ref{fig4.5} illustrates the process of a single neural activation, with reference to the highlighted path in Fig.~\ref{fig4}. Each cycle is divided into two stages: probabilistic activation and self-inhibition. In the first step, read the currently activated neuron, which is the $\mathrm{N}_{i+1}$ neuron in Fig.~\ref{fig4}, indicating that the walker is at position $\mathrm{X}_{i+1}$ in the discrete one-dimensional space. The read result is then scaled by the corresponding weight stored in the synapse $\mathrm{S}_{i+1}$. Subsequently, an activation attempt is made in neurons $\mathrm{N}_{i}$ and $\mathrm{N}_{i+2}$, allowing the walker to move left or right according to the transition probability. During this attempt, the winner-takes-all mechanism ensures that only one neuron is activated. In the second step, after an activation attempt, if successful, the original neuron resets itself in a process known as self-inhibition. Repetition of this process iteratively enables propagation of neural activation through the circuit, simulating random walk patterns of particles.

\subsubsection{Synapse Design}

\begin{figure}[t]
\vspace{-10pt} 
\centering
\includegraphics[width=7cm]{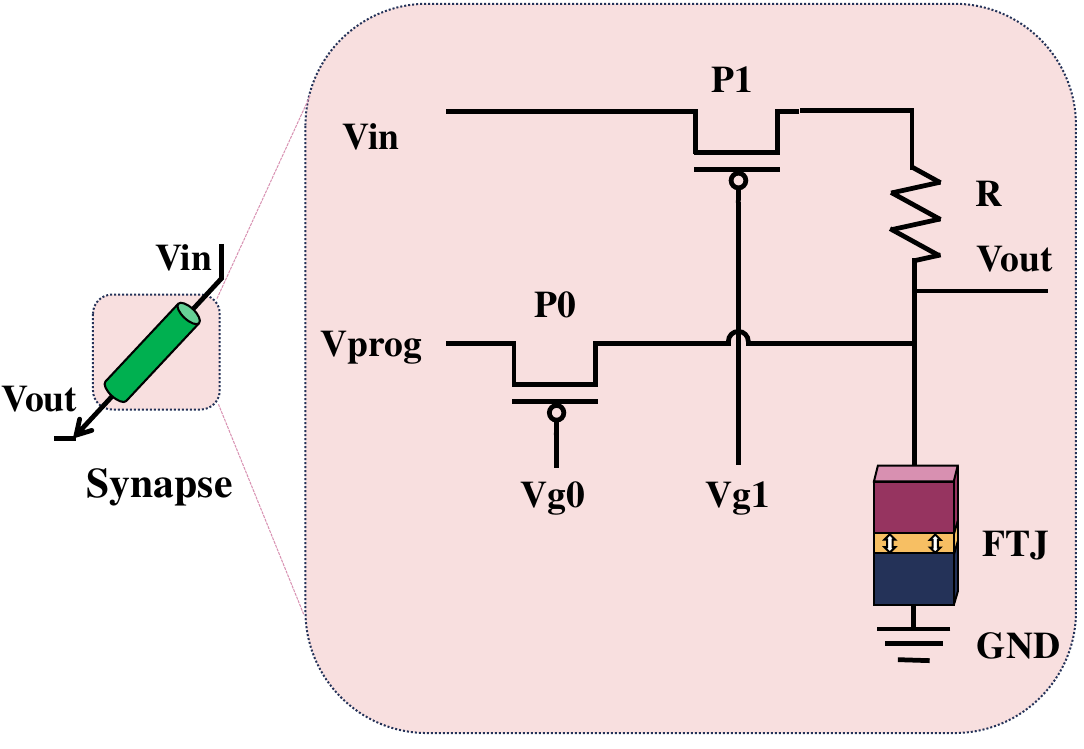}
\caption{DTU synaptic design: leveraging FTJ resistive switching to program synaptic weights.} 
\label{fig6}
\vspace{-10pt} 
\end{figure}

Fig.~\ref{fig6} illustrates an FTJ-based synapse, where $\mathrm{Vin}$ and $\mathrm{Vout}$ serve as input and output pins, respectively, linked to the $\mathrm{rd}$ pin of the presynaptic neuron and the $\mathrm{wr}$ pin of the postsynaptic neuron, facilitating the reception of high voltage output from activated presynaptic neuron and attempting to write to two neighboring neurons. The synapse has two modes of operation: programming and operational. In programming mode, a low gate signal ($\mathrm{Vg0}$) opens transistor P0, allowing $\mathrm{Vprog}$ to be applied to the FTJ and programming its resistive state. In operational mode, a low gate signal ($\mathrm{Vg1}$) opens transistor P1, and the output voltage is determined by the resistive division between $\mathrm{R_{FTJ}}$ and a fixed resistor $\mathrm{R}$:
\begin{equation}
\mathrm{Vout} = \frac{\mathrm{R_{FTJ}}}{\mathrm{R_{FTJ}} + \mathrm{R}} \times \mathrm{Vin}.
\label{equSyn}
\end{equation}

This operation effectively scales $\mathrm{Vin}$ by a weight factor determined by $\mathrm{R_{FTJ}}$. The resulting $\mathrm{Vout}$ then influences the MTJ current, controlling its switching probability. To ensure stable polarization of the FTJ, $\mathrm{Vin}$ is maintained significantly lower than the programming voltage $\mathrm{Vprog}$.

\subsubsection{Neuron Design}
\begin{figure*}[t]
\vspace{-15pt} 
\centering
\includegraphics[width=17cm]{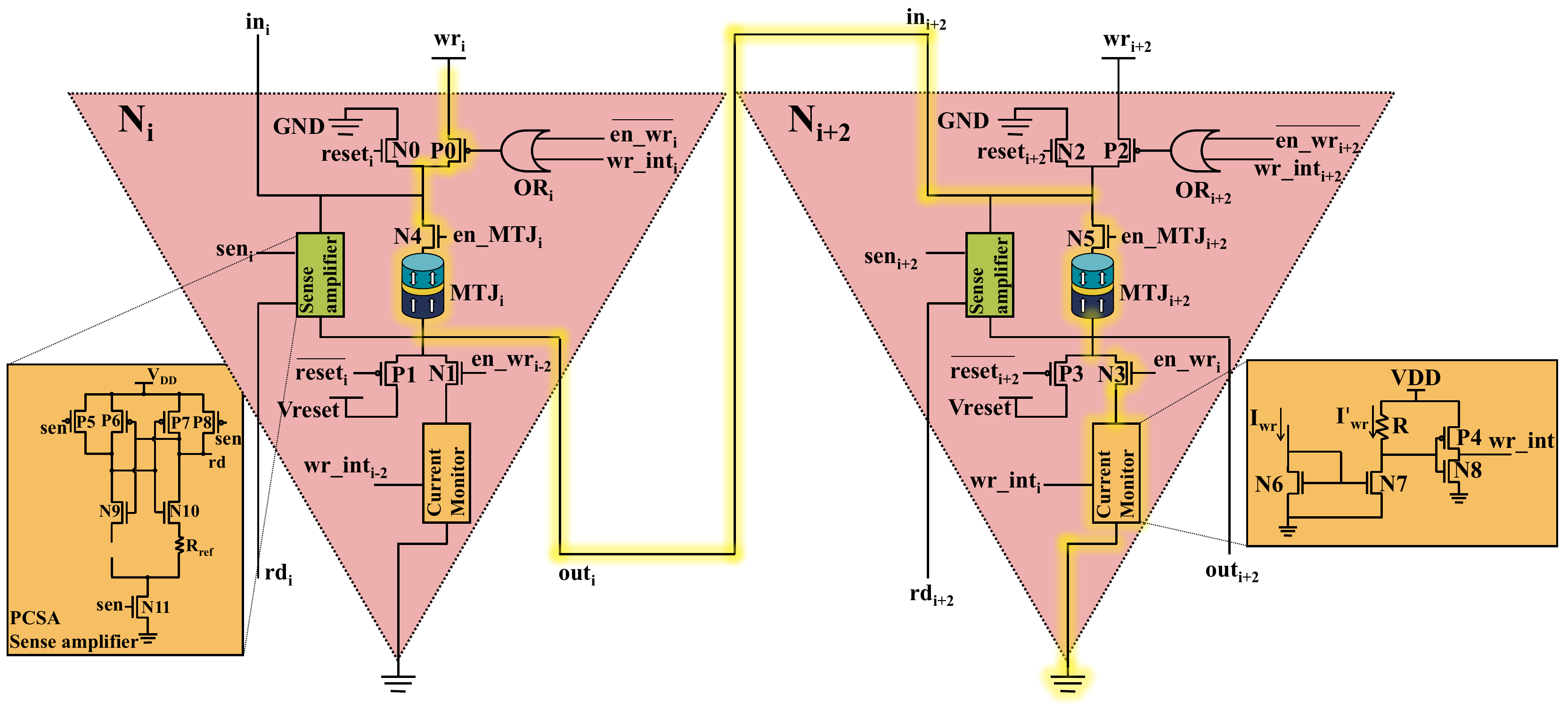}
\caption{Architecture of the DTU neuron, which integrates an MTJ for probabilistic activation, Sense amplifiers for MTJ state detection, a Current monitor to monitor MTJ switching, and supplementary read/write control circuits.} 
\label{fig7}
\vspace{-15pt} 
\end{figure*}

Fig.~\ref{fig7} shows the circuit designs of neurons, with pin arrangements consistent with those in Fig.~\ref{fig4}, focusing on illustrating their workings through examples of the left-neighboring neuron $\mathrm{N}_{i}$ and the right-neighboring neuron $\mathrm{N}_{i+2}$. The functionality of the neuron is detailed according to the activation sequence.

\textbf{Probabilistic Activation:} The operation is initiated when the $\mathrm{sen}$ signal activates the sense amplifier \cite{zhao2009high}, which detects neuronal states and transmits the MTJ status via $\mathrm{rd}$. The PCSA sense amplifier operates in two phases. In the precharge phase, transistors P5, P8, N9, and N10 turn on, equalizing the voltages at the top terminals of the MTJ and reference resistor. In the amplification phase, triggered by the $\mathrm{sen}$ signal, transistor N11 turns on. Due to the resistance difference between the MTJ and the reference resistor, the $\mathrm{rd}$ and its complement are rapidly amplified to high and low levels, completing the readout. Active neurons generate a high voltage output to their synaptic targets. This synaptic output is then directed to $\mathrm{wr}_{i}$. The activation of the P0 transistor is governed by the OR combination of $\mathrm{\overline{en\_wr}}_{i}$ and $\mathrm{wr\_int}_{i}$, requiring both signals to be low for activation. Simultaneously, transistors N3, N4, and N5 become active, allowing the $\mathrm{wr}_{i}$ signal to propagate through $\mathrm{MTJ}_{i}$ to $\mathrm{out}_{i}$. Subsequently, this signal reaches the port $\mathrm{in}_{i+2}$ of the adjacent neuron $\mathrm{N}_{i+2}$, traversing $\mathrm{MTJ}_{i+2}$ before terminating at GND through the Current Monitor (as highlighted in the signal path). This configuration effectively implements a stochastic write operation across the two serially connected MTJ devices.

\textbf{Winner-Takes-All:} The MTJ state transition induces resistance variations, generating current fluctuations in the $\mathrm{wr}_{i}$-$\mathrm{MTJ}_{i}$-$\mathrm{out}_{i}$-$\mathrm{in}_{i+2}$-$\mathrm{MTJ}_{i+2}$-GND path. The Current Monitor circuit \cite{qu2018variation}, composed of a current mirror and an inverter pair, detects these current changes. Upon detection, it asserts $\mathrm{wr\_int}_{i}$ to logic high, which gates off transistor P0 and consequently terminates the write operation.

\textbf{Self-Inhibition:} Neural self-inhibition is achieved through a controlled write process. The signal $\mathrm{wr\_int}_{i}$ regulates the signal $\mathrm{reset}$, which subsequently resets the MTJ of the initially activated neuron.

\vspace{-15pt}
\subsection{Scattering Tracking Unit Design}
\label{STUD}
The DTU efficiently implements fixed-step, discrete-direction random walks such as grid-based Brownian motion in hardware. However, this design exhibits inherent limitations in solving more general PDE problems. Specifically, the random walk mechanism in DTU cannot readily handle complex sampling processes that involve multiple stochastic events, for example, direction-dependent scattering in particle transport. To address these limitations, we introduce the STU. This unit extends the scope of hardware-accelerated PDE solutions by supporting sampling from arbitrary probability distributions.

\subsubsection{Design Philosophy}

\begin{figure}[t]
\vspace{-10pt}
\centering
\includegraphics[width=0.98\linewidth]{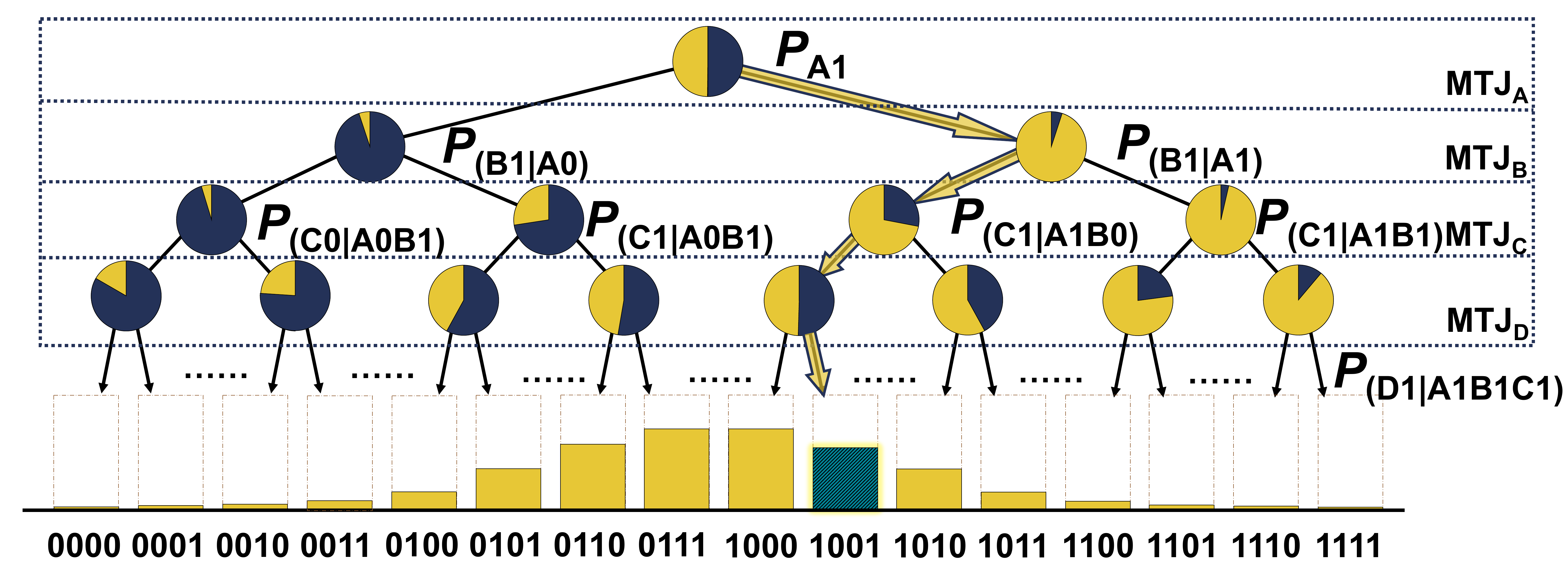}
\caption{MTJ-based four-level conditional probability tree enabling target PDF random numbers.} 
\label{figtree}
\vspace{-15pt}
\end{figure}

The STU tracks stochastic events employing a conditional probability tree-based random number generator, which samples from specific distributions. The conditional probability tree is constructed from multiple stochastic bit generators. In this paper, a configuration of four MTJs ($\mathrm{MTJ_A}$ to $\mathrm{MTJ_D}$) is employed as an illustrative example. 

Fig.~\ref{figtree} depicts the structure of the conditional probability tree and the process of generating the target \textit{probability density function} (PDF). In this structure, each level of the tree corresponds to an individual MTJ, with nodes representing probabilistic MTJ switching. The right branches output ``1" (probability labeled), while the left branches output ``0". The highlighted path exemplifies a random number generation process: $\mathrm{MTJ_A}$ first generates the most significant bit as ``1" with probability $P_{\mathrm{A1}}$, which then conditions $\mathrm{MTJ_B}$ to output ``0" with probability $1-P_{(\mathrm{B1\mid A1})}$. This ``0" subsequently leads $\mathrm{MTJ_C}$ to produce ``0" with probability $1-P_{(\mathrm{C1\mid A1B0} )}$, resulting in a third-bit output of ``0". Given the current three-bit sequence ``100", $\mathrm{MTJ_D}$ generates the least significant bit as ``1" with probability $P_{(\mathrm{D1\mid A1B0C0} )}$, yielding the final output of ``1001". According to the chain rule of conditional probability, the joint probability of obtaining this specific random number is given by:
\vspace{-8pt}
\begin{equation}
\begin{split}
P_{(\mathrm{1001} )} & =  P_{(\mathrm{A1B0C0D1} )}\\ & =  P_{\mathrm{A1}}*(1-P_{(\mathrm{B1\mid A1} )})\\&*(1-P_{(\mathrm{C1\mid A1B0} )})*P_{(\mathrm{D1\mid A1B0C0})}.
\end{split}
\label{pcal}
\end{equation}

The generation probabilities for other 4-bit random numbers are computed using the same method. Thus, each MTJ requires storing 1, 2, 4, and 8 conditional probability values, respectively, selected based on higher-bit outcomes. The STU stores conditional probabilities as synaptic weights. Higher-bit outputs are used to select the corresponding synapses to perform probabilistic MTJ switching. The hardware design directly generates event sampling results using locally stored probability distributions to support complex random walk tracking.

\begin{figure*}[t]
\vspace{-15pt}
\centering
\includegraphics[width=0.9\linewidth]{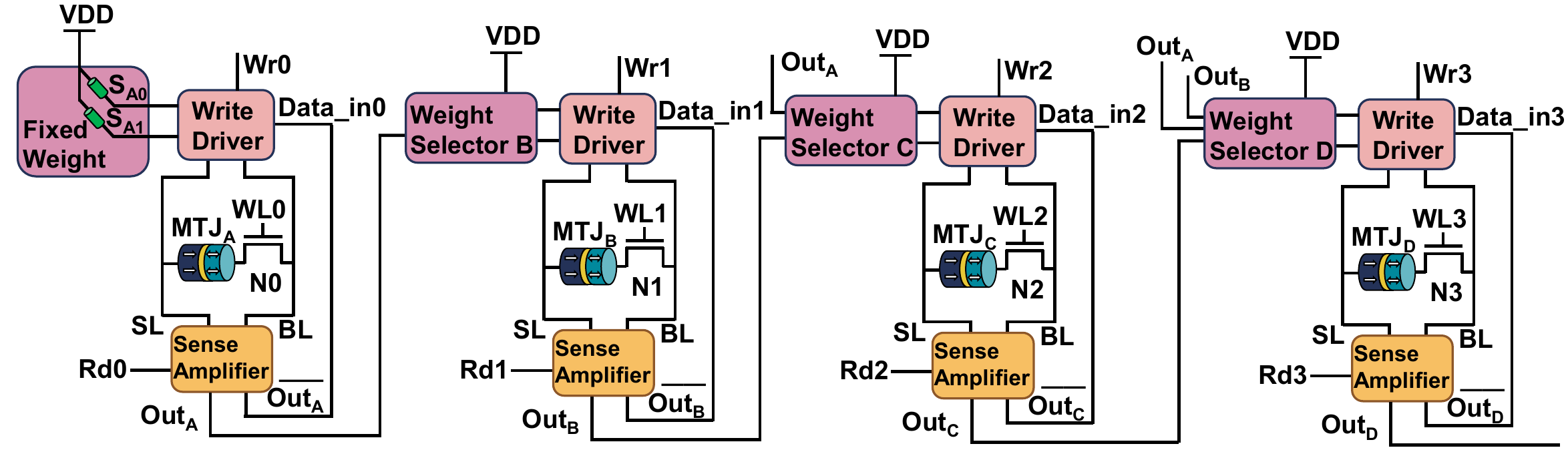}
\caption{Circuit design of STU.} 
\label{figcircuit}
\vspace{-5pt}
\end{figure*}

\begin{figure*}[t]
\centering
\includegraphics[width=0.7\linewidth]{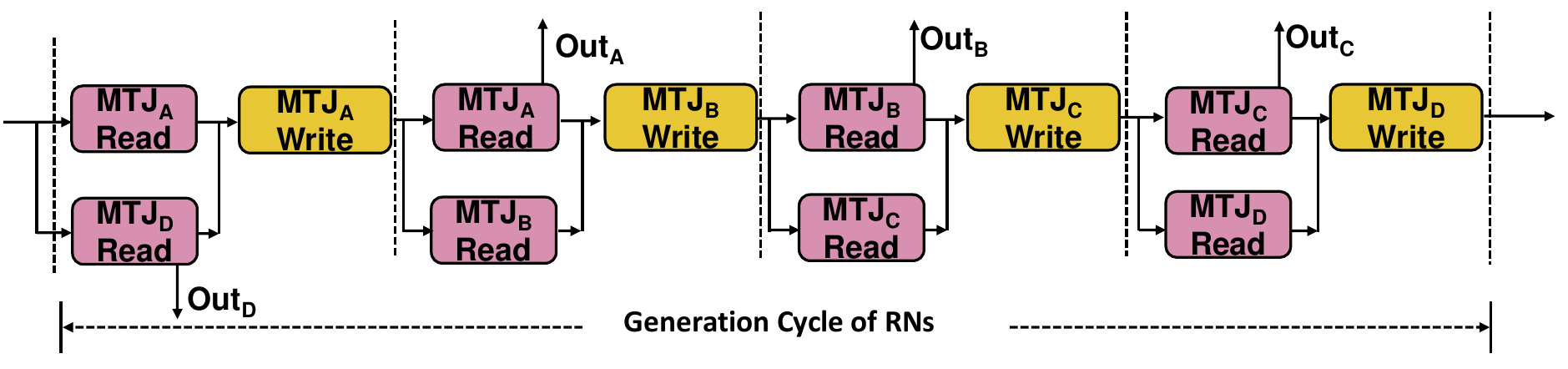}
\caption{Workflow of STU.} 
\label{figlogic}
\vspace{-10pt}
\end{figure*}

\subsubsection{Architecture and Workflow}

As shown in Fig.~\ref{figcircuit}, the STU comprises four similar random bit generation units connected in series. Each unit includes an MTJ, a PCSA sense amplifier \cite{zhao2009high}, and a write driver \cite{Fu2023}, with a weight selector that connects the upper and lower levels. The design and operation of the weight selector will be detailed later.

We use $\mathrm{MTJ_A}$ as an example to illustrate the generation of a single random bit. The random number generation cycle, as illustrated in Fig.~\ref{figlogic}, begins with reading $\mathrm{MTJ_A}$. The sense amplifier detects the state of $\mathrm{MTJ_A}$ and feeds the inverted value of the readout, $\overline{\mathrm{Out_A}}$, into $\mathrm{Data\_in0}$ of the write driver. The write driver then performs a probabilistic write operation on $\mathrm{MTJ_A}$,  with the target state set to $\overline{\mathrm{Out_A}}$. For $\mathrm{MTJ_A}$, the ferroelectric synapses $\mathrm{S_{A0}}$ and $\mathrm{S_{A1}}$ control the probabilities of switching from the AP state to the P state and from P state to AP state, respectively, thus determining the probability that $\mathrm{MTJ_A}$ outputs a ``1". 

Subsequently, the system proceeds to the second phase shown in Fig.~\ref{figlogic}, where it reads $\mathrm{MTJ_A}$ and $\mathrm{MTJ_B}$ simultaneously. The value of $\mathrm{MTJ_A}$ determines the write probability for $\mathrm{MTJ_B}$ ($\mathit{P}_\mathrm{{({B1} \mid {A1})}}$ or $\mathit{P}_\mathrm{{({B1} \mid {A0})}}$ in Fig.~\ref{figtree}), while $\mathrm{MTJ_B}$ writes the inverse of its current state. The third and fourth phases differ only in the configuration of the weight selector. As shown in Fig.~\ref{figtree}, $\mathrm{MTJ_C}$ has four possible output probabilities, determined by two higher-order bits, while the eight output probabilities of $\mathrm{MTJ_D}$ are determined by the first three bits. This approach facilitates the tracking of stochastic events with distribution-matching sampling capabilities, while the architecture remains scalable through the expansion of the weight selector circuitry.

\subsubsection{Weight Selector Design}

\begin{figure}[t]
\vspace{-10pt}
    \centering
    \begin{minipage}[b]{0.4\linewidth}
        \centering
        \includegraphics[width=\linewidth]{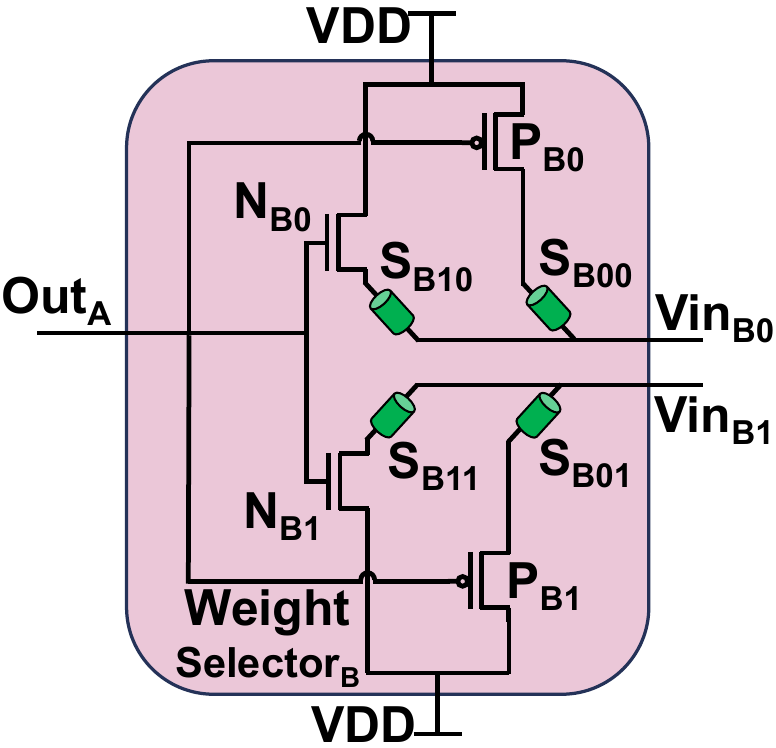}
        \centerline{(a) Weight selector B.}
        \label{subfig:b}
    \end{minipage}
    \hfill
    \begin{minipage}[b]{0.58\linewidth}
        \centering
        \includegraphics[width=\linewidth]{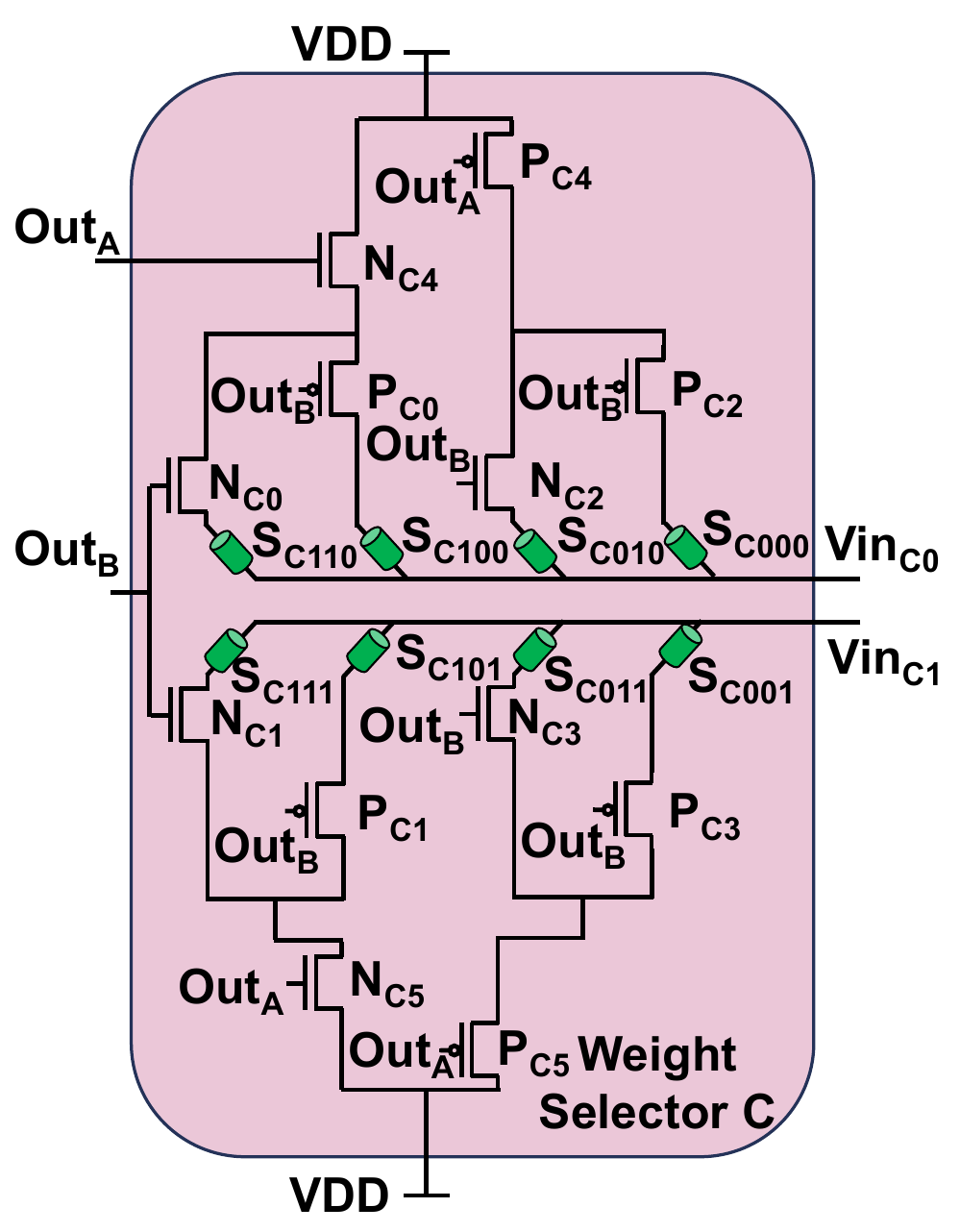}
        \centerline{(b) Weight selector C.}
        \label{subfig:c}
    \end{minipage}
    
    \vspace{-10pt}
    \begin{minipage}[b]{\linewidth}
        \centering
        \includegraphics[width=0.9\linewidth]{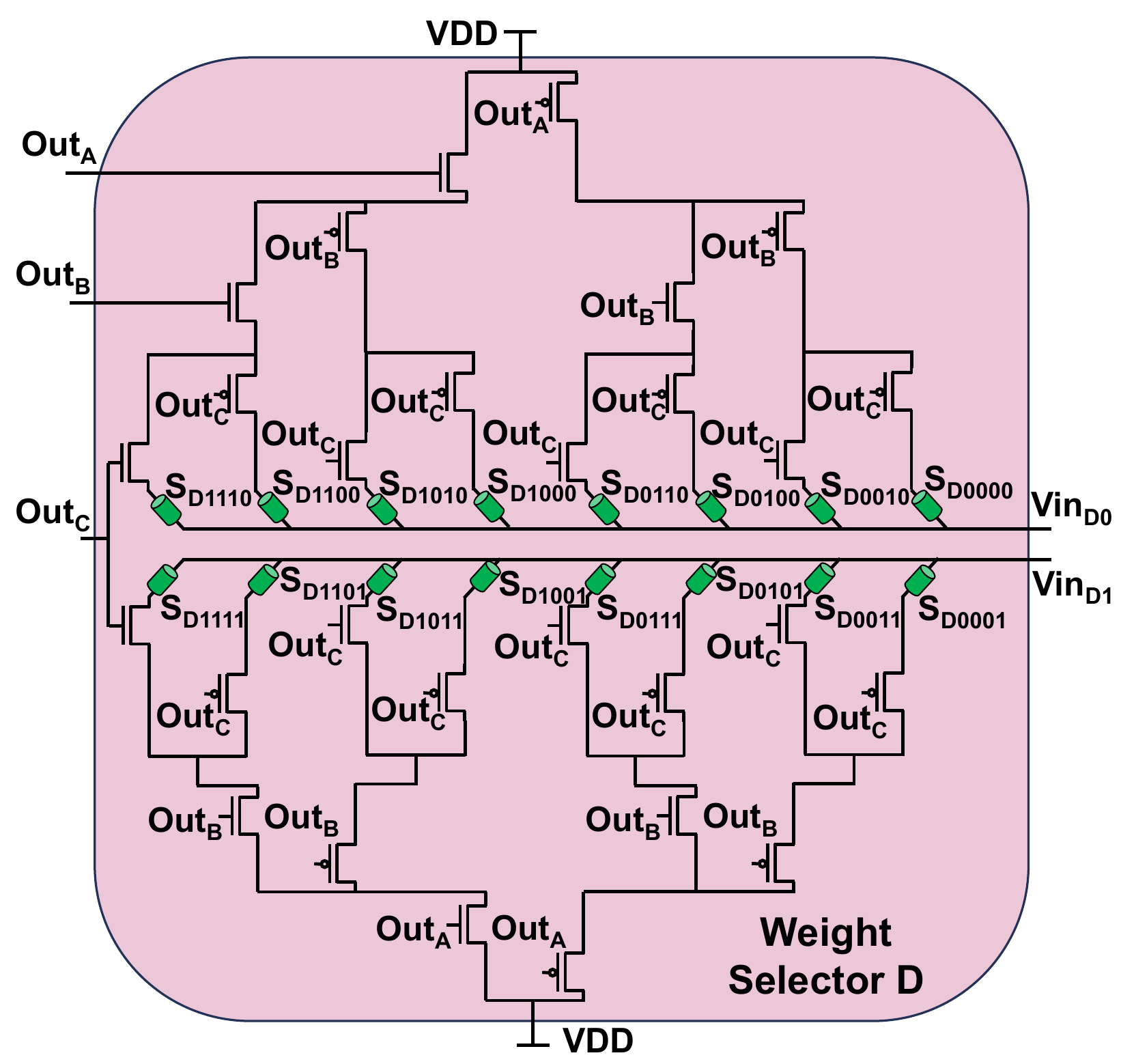}
        \centerline{(c) Weight selector D.}
        \label{subfig:d}
    \end{minipage}
    \caption{Structure of weight selectors.}
    \label{fig:selectors}
    \vspace{-10pt}
\end{figure}

The weight selector chooses the synapse pair that stores the corresponding probability weight based on the state of the higher-level MTJ.

Fig.~\ref{fig:selectors}(a) shows the structure of the weight selector B, which serves as the foundation for subsequent weight selectors. The circuit input, $\mathrm{Out_A}$, represents the $\mathrm{MTJ_A}$ value read at the beginning of the second stage in Fig.~\ref{figlogic}. If $\mathrm{Out_A}$ is ``1", transistors $\mathrm{N_{B0}}$ and $\mathrm{N_{B1}}$ are turned on, $\mathrm{P_{B0}}$ and $\mathrm{P_{B1}}$ are turned off, and $\mathrm{VDD}$ is applied to $\mathrm{S_{B10}}$ and $\mathrm{S_{B11}}$, which store the probabilities to write ``0" or ``1" to $\mathrm{MTJ_B}$. $\mathrm{Vin_{B0}}$ and $\mathrm{Vin_{B1}}$ are then sent to $\mathrm{Vin_{0}}$ and $\mathrm{Vin_{1}}$ of the write driver, controlling the $\mathrm{MTJ_B}$ write operation. When $\mathrm{Out_A}$ is ``0", the synapses $\mathrm{S_{B00}}$ and $\mathrm{S_{B01}}$ are activated instead.

Fig.~\ref{fig:selectors}(b) shows the structure of the weight selector C. For $\mathrm{MTJ_C}$, four output combinations of $\mathrm{Out_A}$ and $\mathrm{Out_B}$ correspond to the selection of four pairs of synapses. When both $\mathrm{Out_A}$ and $\mathrm{Out_B}$ are ``1", transistors $\mathrm{N_{C0}}$, $\mathrm{N_{C1}}$, $\mathrm{N_{C4}}$, and $\mathrm{N_{C5}}$ are turned on, with $\mathrm{VDD}$ enabling pairs of synapses $\mathrm{S_{C110}}$ and $\mathrm{S_{C111}}$. The selection conditions for other pairs of synapses follow the same principle, with the three bits in the synapse name representing the values of $\mathrm{Out_A}$, $\mathrm{Out_B}$, and the writing direction of $\mathrm{MTJ_C}$.

Fig.~\ref{fig:selectors}(c) shows the structure of the weight selector D. $\mathrm{MTJ_D}$ has eight selectable synapse pairs, with the naming convention reflecting the values of $\mathrm{Out_A}$, $\mathrm{Out_B}$, and $\mathrm{Out_C}$, as well as the write direction to $\mathrm{MTJ_D}$.

Notably, the number of selectable synapse pairs grows exponentially with the tree depth (weight selector B: 2 pairs, C: 4 pairs, D: 8 pairs), reflecting the $O(2^n)$ hardware complexity inherent to representing an arbitrary distribution over $2^n$ discrete states. This scaling is a mathematical property of the conditional probability tree rather than a circuit-level deficiency, and it imposes a practical trade-off between discretization resolution and hardware cost. For MC-based PDE problems with moderate angular resolution requirements, a small $n$ (e.g., 4--5 bits) is sufficient to achieve acceptable accuracy, as demonstrated in our experiments; for applications demanding substantially finer discretization, alternative discretization or heterogeneous computing strategies should be considered.

\section{Experiments and Evaluation}
\label{Experiments}

\subsection{Experimental Setup}

In the experimental section, we evaluate both the circuit functionality of the DTU and STU, as well as their capability to perform stochastic tracing and solve PDEs. SPICE-level simulations were performed using \textit{\SI{45}{\nano\meter} generic process development kit} (GPDK045), integrating compact MTJ \cite{Wu2022} and FTJ \cite{wang2014compact} models. The MC simulations incorporate  MTJ stochasticity and $3\sigma$ process variations, with key parameters listed in Table~\ref{tab1}, with FTJ parameters listed in Table~\ref{tabftj}.  All simulations were performed at a temperature of 27$^\circ \text{C}$ (\SI{300}{K}). The software baseline was implemented in Python 3.7 on an Intel\textsuperscript{\textregistered} Core\texttrademark\ i9-12900 CPU (Ubuntu 20.04.1), using the built-in {\tt random} library for RNG and the {\tt cachetools} library for cache modeling.

\begin{table}[t]
\centering
\vspace{-10pt}
\caption{Key device parameters for MTJ compact model.}\label{tab1}
\resizebox{0.45\textwidth}{!}{
\renewcommand{\arraystretch}{1.1} 
\begin{tabular}{
>{}c |
>{}c |
>{}c }
\hline
{ \textbf{Parameter}} & {\textbf{Description}}  & {\textbf{Value}}  \\ \hline \hline
{$t_{\mathrm{FL}}$}       & {Thickness of the free layer}        & {1.3nm}  \\ \hline
{$\sigma _{t_{\mathrm{FL}}} $}       & {Standard deviation of $t_{\mathrm{FL}}$}        & {3\% of 1.3nm}  \\ \hline

{$CD$}         & {Critical diameter}        & {32nm}   \\ \hline
{$t_{\mathrm{TB}}$}       & {Thickness of the tunnel barrier}     & {0.85nm} \\ \hline

{$\sigma _{t_{\mathrm{TB}}} $}       & {Standard deviation of $t_{\mathrm{TB}}$}     & {3\% of 0.85nm} \\ \hline

{$TMR$}       & {TMR ratio} & {200\%}   \\ \hline
{$\sigma _{TMR}$}       & {Standard deviation of TMR} & {3\% of 200\%}   \\ \hline
\end{tabular}
}
\end{table}

\begin{table}[t]
\centering
\vspace{-5pt}
\caption{Key device parameters for FTJ compact model.}\label{tabftj}
\resizebox{0.5\textwidth}{!}{
\renewcommand{\arraystretch}{1.1} 
\begin{tabular}{
>{}c |
>{}c |
>{}c }
\hline
{ \textbf{Parameter}} & {\textbf{Description}}  & {\textbf{Value}}  \\ \hline \hline
{$t_{B}$}       & {Barrier thickness}        & {2nm}  \\ \hline
{$r$}        & {Junction surface radius}  & {175nm}   \\ \hline
{$U_{\mathrm{N}}$} & {Creep energy barrier for the domain} & {0.67eV} \\ \hline
{$U_{\mathrm{P}}$} & {Creep energy barrier for the domain wall} & {0.52eV} \\ \hline
{$\tau_{\mathrm{0N}}$} & {Attempt time of the domain nucleation} & {2.8e-15s} \\ \hline
{$\tau_{\mathrm{0P}}$} & {Attempt time of the domain wall} & {9e-14s} \\ \hline
{$\varphi_{\mathrm{1OFF}}$} & {Barrier potential height at
LSMO/BTO interface(OFF)} & {0.678V} \\ \hline
{$\varphi_{\mathrm{1ON}}$} & {Barrier potential height at
LSMO/BTO interface(ON)} & {0.53V} \\ \hline
{$\varphi_{\mathrm{2OFF}}$} & {Barrier potential height at
Co/BTO interface(OFF)} & {0.978V} \\ \hline
{$\varphi_{\mathrm{2ON}}$} & {Barrier potential height at
Co/BTO interface(ON)} & {1.014V} \\ \hline
{$m_{\mathrm{OFF}}$} & {Effective electron mass(OFF)} & {0.931me} \\ \hline
{$m_{\mathrm{ON}}$} & {Effective electron mass(ON)} & {0.437me} \\ \hline

\end{tabular}
}
\vspace{-15pt} 
\end{table}

\vspace{-10pt}
\subsection{Diffusion Tracking Unit Simulations}

\begin{figure}[t]
\centering
\includegraphics[width=8.5cm]{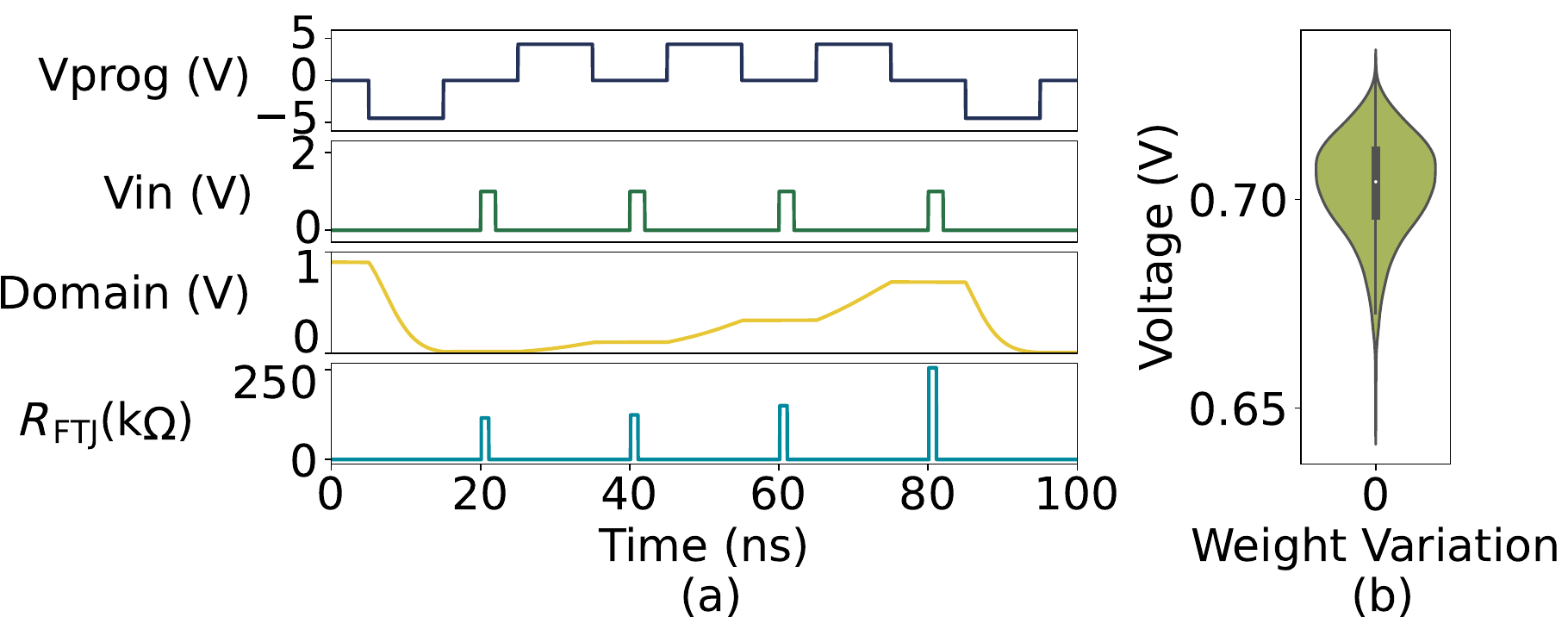}
\vspace{-5pt}
\caption{(a) Transient simulation results for ferroelectric synapses, (b) MC simulation results for ferroelectric synapses.}
\label{figsy}
\vspace{-15pt} 
\end{figure}

\subsubsection{Circuit-Level Simulation}
The circuit-level simulation of DTU includes the evaluation of ferroelectric synapses for weight storage functionality, as well as the evaluation of neuronal characteristics including probabilistic activation, winner-takes-all behavior, and self-inhibition mechanisms.

Fig.~\ref{figsy}(a) illustrates the alternating operation of the synapse in programming and write-driving modes. From top to bottom, the waveforms correspond to $\mathrm{Vprog}$, $\mathrm{Vin}$, the direction of the FTJ domain wall, and the FTJ resistance, as shown in Fig.~\ref{fig6}. Between \SI{5}{\nano\second}-\SI{15}{\nano\second}, $\mathrm{Vprog}$ applies a negative voltage to program the FTJ, resetting the domain wall direction to 0. At \SI{20}{\nano\second}, the synapse is in write-driving mode, where $\mathrm{Vin}$ applies a low voltage to read the FTJ resistance. During the intervals \SI{25}{\nano\second}-\SI{35}{\nano\second}, \SI{45}{\nano\second}-\SI{55}{\nano\second}, and \SI{65}{\nano\second}-\SI{75}{\nano\second}, $\mathrm{Vprog}$ applies the programming voltage, inducing the growth of the domain wall. The programming gaps are driven by $\mathrm{Vin}$, and the FTJ resistance increases as the domain wall evolves. Finally, between \SI{85}{\nano\second}-\SI{95}{\nano\second}, a negative programming voltage resets the FTJ.

Fig.~\ref{figsy}(b) presents the MC simulation results for a \SI{15}{\nano\second} programming voltage applied to a reset synapse, illustrating the output voltage distribution under process variations. After 50,000 simulations, the mean change is less than $0.32\%$, with a variance below 0.000138. This variation corresponds to an effective resolution of approximately 8 bits ($1/2^8 = 0.3906\%$). This bounded variability is incorporated into our system-level simulations to assess computational fidelity under the limited precision of FTJ-based synapses.

\begin{figure}[t]
\centering
\includegraphics[width=9cm]{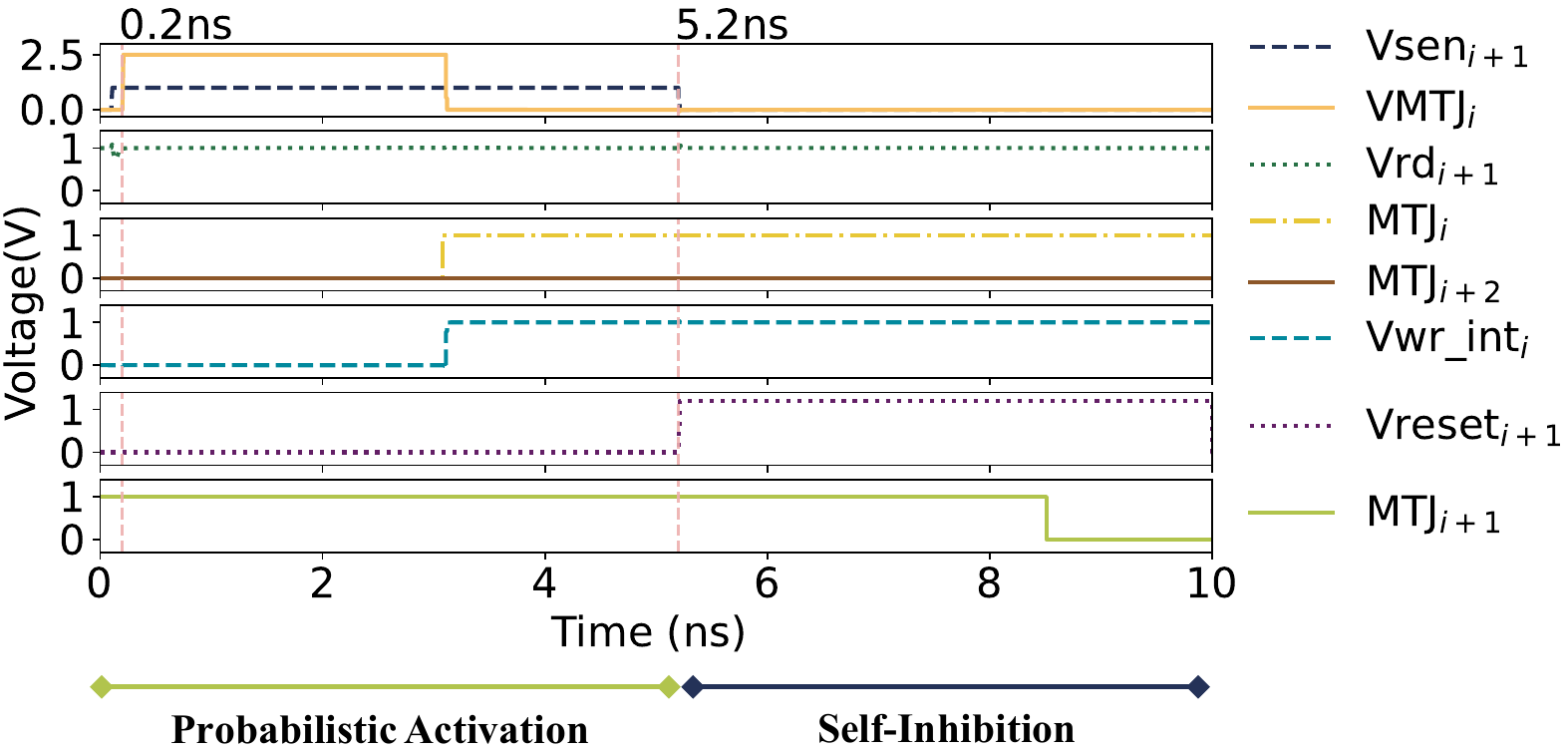}
\caption{The transient simulates probabilistic activation of neurons $\mathrm{N}_{i}$ and $\mathrm{N}_{i+2}$ by $\mathrm{N}_{i+1}$, with the figure illustrating the activation process of neuron $\mathrm{N}_{i}$.} 
\vspace{-15pt} 
\label{figneu}
\end{figure}

Fig.~\ref{figneu} shows a neuronal activation event within a cluster involving a presynaptic neuron $\mathrm{N}_{i+1}$ and postsynaptic neurons $\mathrm{N}_{i}$ and $\mathrm{N}_{i+2}$. Activation pathways are highlighted in Fig.~\ref{fig4}, with details of the neuronal circuits provided in Fig.~\ref{fig7}. Activation begins when the signal $\mathrm{Vsen}_{i+1}$ triggers the sense amplifier in the neuron $\mathrm{N}_{i+1}$, which completes pre-charging and reading within \SI{0.2}{\nano\second} and transmits a logical ``1" via $\mathrm{Vrd}_{i+1}$. This weighted signal propagates through the synapses to become $\mathrm{Vwr}_{i}$ for the neuron $\mathrm{N}_{i}$. Within the following \SI{5}{\nano\second}, $\mathrm{Vwr}_{i}$ is applied through transistor P0 to the source of N4, generating $\mathrm{VMTJ}_{i}$ to attempt writing on both $\mathrm{MTJ}_{i}$ and $\mathrm{MTJ}_{i+2}$. In this simulation, $\mathrm{MTJ}_{i}$ switches states, resulting in increased resistance and reduced current. The Current Monitor in the neuron $\mathrm{N}_{i+2}$ responds by increasing its output ($\mathrm{Vwr\_int}_{i}$) and pulling down $\mathrm{VMTJ}_{i}$, thus maintaining $\mathrm{MTJ}_{i+2}$ unchanged and implementing the winner-takes-all mechanism. Subsequently, self-inhibition suppresses the activation of the original neuron. The $\mathrm{Vwr\_int}_{i}$ signal registers the activation event and controls $\mathrm{Vreset}_{i+1}$ to output a ``1" during the inhibition phase, ensuring that $\mathrm{MTJ}_{i+1}$ is reset within the required time frame. Additional MC simulations incorporating process variations and MTJ switching stochasticity generate a neuron activation history table from 50,000 simulation runs, which is subsequently employed in system-level simulations.

\subsubsection{System-Level Simulation}

In this section, we explore a steady-state problem with boundary conditions. We present a specific example of a \textit{one-dimensional} (1D) steady-state heat equation: A thin metal wire of length $L$  has one end at $x=0$ exposed to an external temperature $T=0$, and the thermal gradient is zero. At the opposite end, there is a heat source with a gradient of $-F$, which decreases linearly toward the left endpoint. The steady-state temperature distribution along the wire at position $x$, denoted by $u(x)$, is given by:
\begin{equation}
    \begin{split}
   &0 =\frac{\mathrm{d^{2} } }{\mathrm{d} x^{2} } u-F(L-x), x\in [0,L],\\
          &u(0) = 0,  \quad
       {u}'(0) = 0.
    \end{split}
\label{equ2}
\end{equation}

This problem has an analytical solution as follows:
\begin{equation}
u(x)=\frac{FLx^2}{2}- \frac{Fx^3}{6}.
\end{equation}

\begin{figure*}[t]
\centering
\vspace{-10pt} 
\includegraphics[width=13cm]{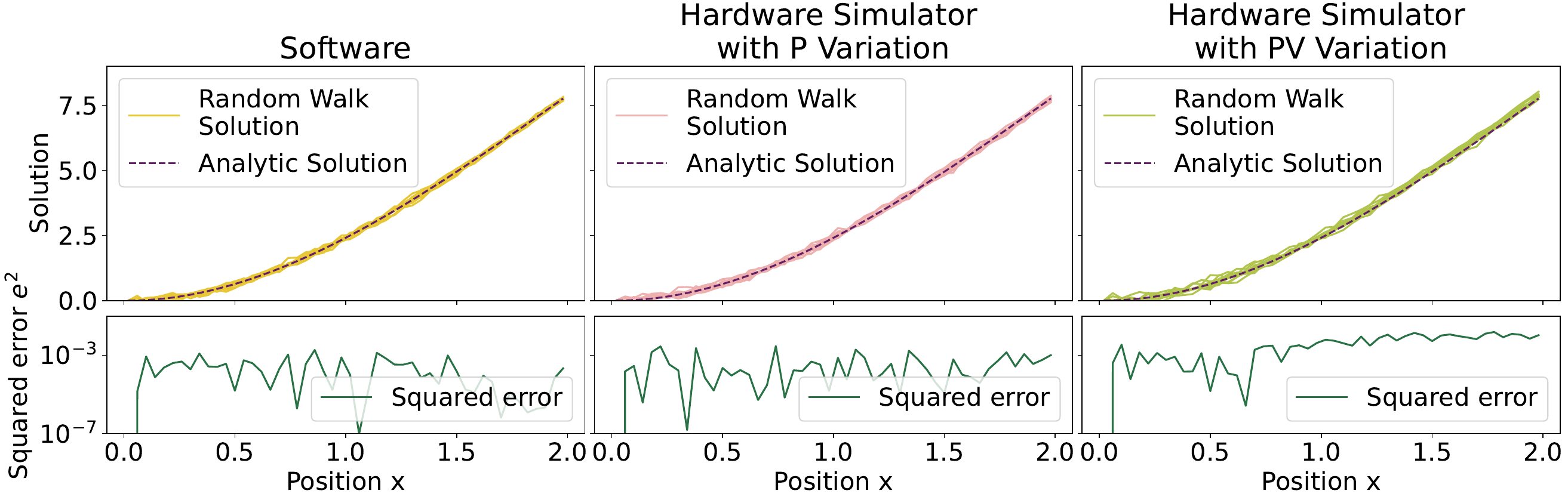}
\caption{Comparison of squared errors in solving the steady-state problem between software-based random walk and its hardware counterpart under the impact of process and voltage variations.
} 
\label{fig13}
\vspace{-10pt} 
\end{figure*}

Next, we construct a Markov chain to execute random walks to solve this equation using the MC random walk method. Detailed information can be found in this work \cite{smith2020solving}.

\begin{itemize}
    \item A 1D space of length $L$ is discretized into $N$ positions, with $W$ walkers initialized at each position.
    \item Simulate particles among these $W$ walkers: Each step takes a time $dt$. The particle has a probability $P_{s}$ of staying in the same position and a probability $P_{g}$ of moving left or right. In position $x=0$, the probability of moving right is $2P_{g}$.
    \item Track each walker until it reaches $x=L$ and then stop the simulation. Record the initial position $\mathrm{X}_i$ and count the number of passages through $\mathrm{X}_j$ as $n_{i,j}$, forming a matrix documenting the walking histories.
    \item Calculate the solution to the PDE using Equation (\ref{equux}).
\end{itemize}
\begin{equation}
    \begin{split} 
    &\mathbb{E} [-F\int_{0}^{T}L-X(s)\,\mathrm{d}s  \mid X_{0} = \mathrm{X}_{i}] \\
    &\approx -\frac{F\cdot dt}{W}\sum_{j}n_{i,j}(L-\mathrm{X}_{j}):= u_{i}.\\
    &u(\mathrm{X}_{i}) \approx u_{i}-u_{0}.
    \end{split}
\label{equux}
\end{equation}

The parameter values in our simulations are $L=2$, $N=50$, $dt=0.00038$, $F=3$, and $W=1e4$. Fig.~\ref{fig13} shows three PDE solutions: software-driven random walks, hardware simulations with process (P) variations, and with process and voltage (PV) variations. The voltage variation, applied to $V_{\mathrm{wr}_i}$ in Fig.~\ref{fig7}, originates from the synaptic weight distribution in Fig.~\ref{figsy}(b) and reflects the limited precision of FTJ-based synapses, including synaptic drift and weight-update errors. The dashed line represents the analytical solution ($C_{\mathrm{an}}$), while the 10 iterations of random walk solutions ($C_{\mathrm{rw}}$) using MC random walks are shown nearby, with squared error $e ^{2}$ calculated with the following equation: 
\begin{equation}
e ^{2}_i = \left | \overline{C_{\mathrm{rw}} (\mathrm{X}_{i})}   -C_{\mathrm{an}}(\mathrm{X}_{i})\right | ^{2}.
\label{variation}
\end{equation}

Our DTU design effectively resolves PDEs, with simulated hardware results showing a squared error below 1e-3 when considering process variation, which is consistent with the software-based solution. Introducing voltage offsets from synaptic precision variations slightly increases the squared error as $x$ approaches $L$, due to a slight reduction in the mean synaptic weight distribution from precision and drift. Meanwhile, the non-linear dependence of the walk probability on the activation voltage lowers the expected activation probability, increasing the value of the solution. However, the large number of walks may provide error tolerance, with the squared error remaining below 1e-2, which suggests that the DTU is tolerant to voltage variations.

\vspace{-10pt}
\subsection{Scattering Tracking Unit Simulations}

\subsubsection{Circuit-Level Simulation}

\begin{figure}[t]
\vspace{-10pt}
\centering
\includegraphics[width=0.9\linewidth]{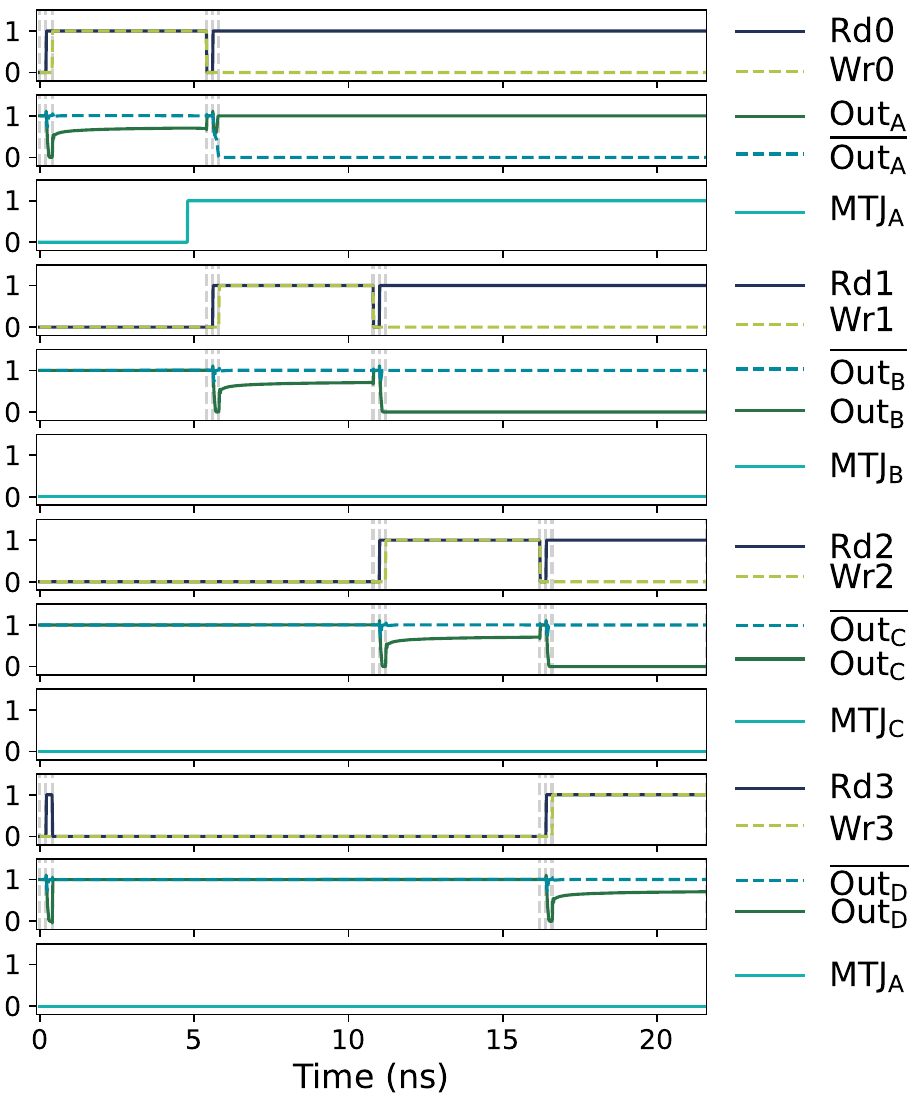}
\caption{Sequential timing control and probabilistic switching in MTJ read/write cycles for STU circuit.} 
\label{fig15}
\vspace{-10pt}
\end{figure}

The circuit-level simulation of the STU primarily evaluates its ability to generate random samples. The simulation results for one random sampling cycle of the STU circuit, governed by the control logic in Fig.~\ref{figlogic}, are shown in Fig.~\ref{fig15}. The MTJ read process consists of a \SI{0.2}{\nano\second} pre-charge sub-phase ($t_{\mathrm{pre}}$) followed by a \SI{0.2}{\nano\second} amplification sub-phase ($t_{\mathrm{rd}}$). The MTJ write time ($t_{\mathrm{wr}}$) is set to \SI{5}{\nano\second}, enabling the output probability to span from $0\%$ to $100\%$ through changes in the write voltage.

The random sampling cycle sequentially processes $\mathrm{MTJ_A}$, $\mathrm{MTJ_B}$, $\mathrm{MTJ_C}$, and $\mathrm{MTJ_D}$, with the write probability of each MTJ determined by the higher-order bits, implementing the conditional probability tree mechanism. $\mathbf{MTJ_A}$: After a \SI{0.2}{\nano\second} precharge, $\mathrm{Rd0}$ reads its state ($\mathrm{Out_A}$), and its complement ($\mathrm{\overline{Out_A}}$) is fed back. From \SI{0.4}{\nano\second} to \SI{5.4}{\nano\second}, $\mathrm{Wr0}$ triggers a probabilistic write, with the result read at \SI{5.4}{\nano\second} - \SI{5.8}{\nano\second} (switching to logic ``1" here). $\mathbf{MTJ_B}$: $\mathrm{Rd1}$ reads at \SI{5.6}{\nano\second}, and $\mathrm{Wr1}$ executes a probabilistic write from \SI{5.8}{\nano\second} - \SI{10.8}{\nano\second}, based on $\mathrm{Out_A}$. $\mathbf{MTJ_C}$: $\mathrm{Rd2}$ reads at \SI{11}{\nano\second}, and $\mathrm{Wr2}$ writes from \SI{11.2}{\nano\second} - \SI{16.2}{\nano\second}, conditioned on the prior $\mathrm{Out_A}$ and $\mathrm{Out_B}$ results. $\mathbf{MTJ_D}$: $\mathrm{Rd3}$ reads at \SI{16.4}{\nano\second}, and $\mathrm{Wr3}$ writes from \SI{16.6}{\nano\second} - \SI{21.6}{\nano\second}. The final read for $\mathrm{MTJ_D}$ overlaps with the first read phase of $\mathrm{MTJ_A}$ in the subsequent cycle.

In the cycle shown in Fig.~\ref{fig15}, the STU starts in the state ``0000" (decimal 0), and the generated result is ``1000" (decimal 8). By performing MC simulations on the $2^4$ possible initial states of the 4-bit sequence and constructing a decimal sequence based on the relationship between pre-generation and post-generation states, we can obtain the random events sequence sampled by the STU in the SPICE simulation.

\subsubsection{System-Level Simulation}

\begin{figure}[t]
\vspace{-10pt}
\centering
\includegraphics[width=0.6\linewidth]{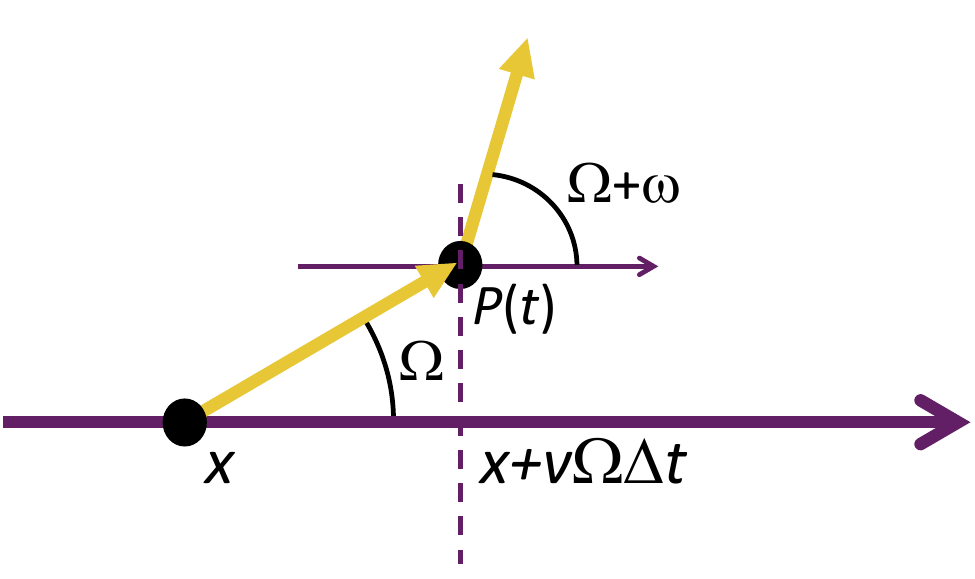}
\caption{Spatial particle transport model, in which particles undergo $P(t)$-governed scattering, altering their direction.} 
\label{fig18}
\vspace{-10pt}
\end{figure}

\begin{figure*}[t]
\vspace{-20pt}
\centering
\includegraphics[width=0.9\linewidth]{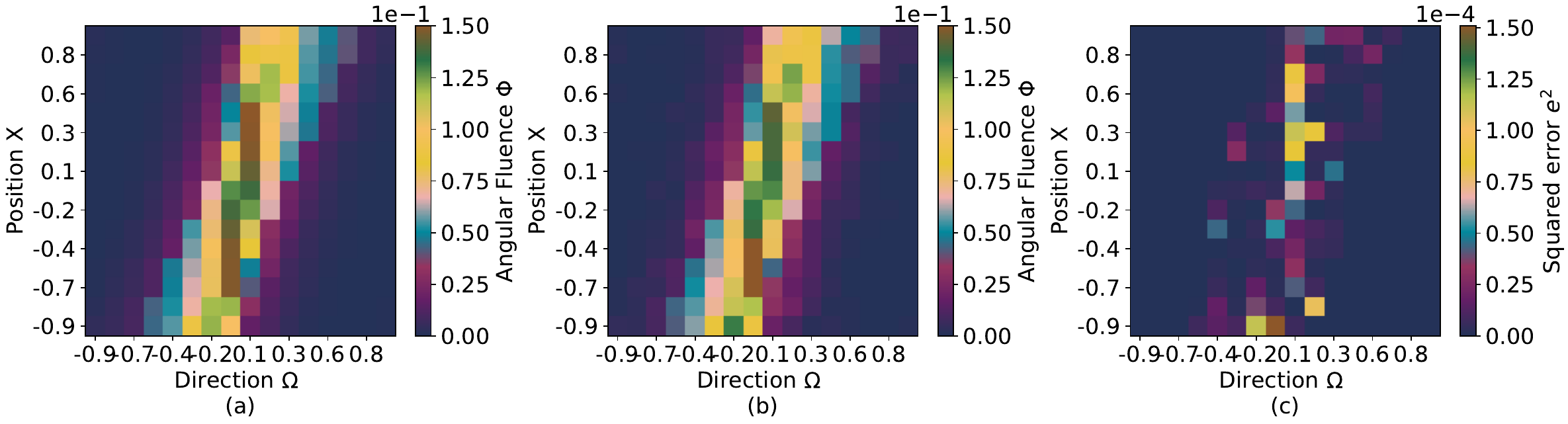}
\caption{MC solutions for particle transport simulations: (a)angular flux density computed on CPU via software method, (b) angular flux density computed via STU, (c) squared errors comparison.} 
\label{fig19}
\end{figure*}

We consider a particle transport problem involving angular flux density. The particle motion in space includes scattering events at a rate $\Sigma _{s}$, which is projected onto a one-dimensional space, as shown in Fig.~\ref{fig18}. The horizontal axis represents position $x$, while the direction of particles is defined by $\Omega$. Scattering occurs probabilistically according to a Poisson process $P(t)$, and after scattering, the new direction is drawn from a Gaussian distribution with a mean aligned with the original direction. We focus on the domain $x\in [-1,1]$ with an absorbing boundary condition, leading to the following problem:

\vspace{-15pt}
\begin{equation}
\begin{split}
0 & = -v\Omega\frac{\partial }{\partial x} \Phi (x,\Omega)+vR(x) +v\Sigma _{s}\int (\Phi (x,\Omega+\omega) \\
&\quad -\Phi (x,\Omega))p^{\ast}(\omega+\Omega\mid \Omega) \mathrm{d}\omega,\\
&\quad x\in (-1,1),\quad \Omega\in [-1,1], \\
\Phi (1,\Omega) & = 0, \quad \text{if} \quad\Omega<0,\\
\Phi (-1,\Omega) & = 0, \quad \text{if} \quad\Omega>0.
\end{split}
\label{equMC}
\vspace{-20pt}
\end{equation}

The particle velocity is $v=200$ and the scattering rate is $\Sigma _{s}=0.5$. The particle source is defined as follows:

\vspace{-8pt}
\begin{equation}
 R(x)=\left\{
\begin{array}{ll}
0.015 & \text{if} \quad \left|x\right|  <0.5 \\
0 & \text{otherwise}.
\end{array}
\right.
\label{equR}
\end{equation}

In the case of Poisson events, scattering occurs, with the new direction after scattering determined by a Gaussian distribution with variance $\sigma ^2 = 4$:
\vspace{-6pt}
\begin{equation}
p(\omega) = \frac{1}{\sqrt{2 \pi \sigma^2}} \exp \left( -\frac{\omega^2}{2 \sigma^2} \right).
\label{equG}
\end{equation}

Due to the inherent complexity of the particle transport problem, an analytical solution is unavailable. Instead, a probabilistic solution is derived through MC simulations that model the motion of particles. By discretizing position $x$ and direction $\Omega$, then constructing a Markov chain through MC-simulated random walks to track particle histories, we obtain probabilistic solutions for angular flux density. The solution method matches Smith et al. \cite{smith2022neuromorphic} but uses a Gaussian distribution for particle scattering, better reflecting real-world conditions.

The MC simulation process, executed on the CPU, yields the angular flux density solution shown in Fig.~\ref{fig19}(a). The x-axis represents particle angle $\Omega$, the y-axis represents particle position $x$, and the color bar indicates $\Phi(x,\Omega)$, with a number of walkers $W=10,000$. Since Equation (\ref{equMC}) lacks an analytical solution, this result is adopted as the approximate solution.

\begin{figure*}[t]
	\centering
	\includegraphics[width=0.9\linewidth]{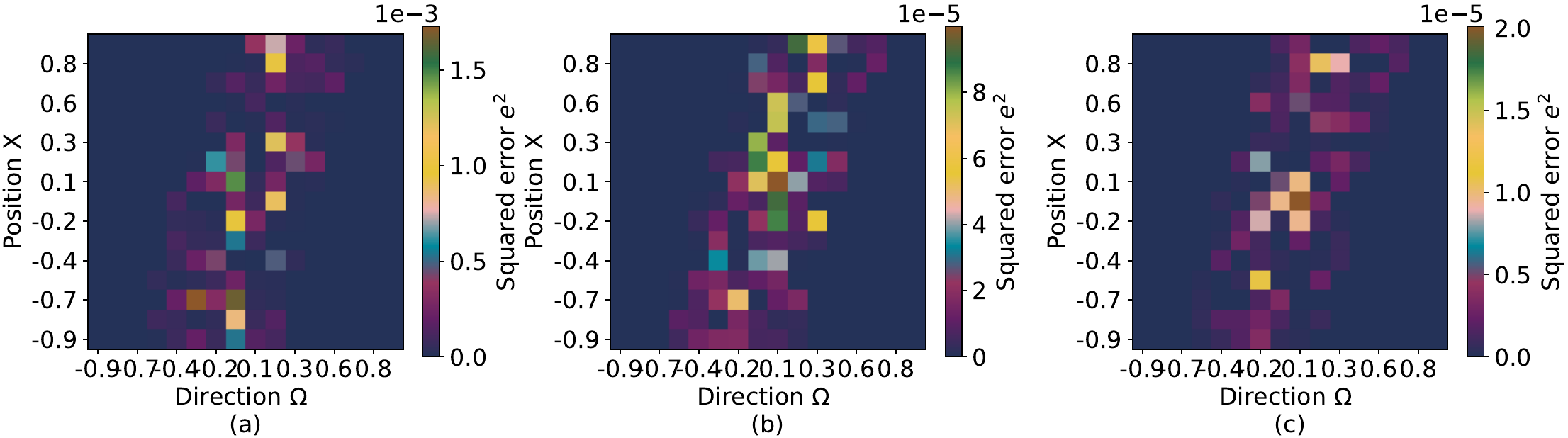}
	\caption{Squared error of Software-Based Particle Transport Solutions for Different Scales with the Result for $W=10,000$: (a) $W=10$, (b) $W=100$, (c) $W=1,000$.} 
	\label{figMCW}
\end{figure*}

\begin{table}[t]
	\centering
	\caption{Particle tracking counts for different simulation scales ($W=10, 100, 1,000$ and $W=10,000$).}\label{tab4}
	\resizebox{0.3\textwidth}{!}{
		\begin{tabular}{
				>{}c |
				>{}c }
			\hline
			{ \textbf{W}} & {\textbf{Particle tracking counts}}    \\ \hline \hline
			{10}      & {3,127}         \\ \hline
			{100}      & {30,337}        \\ \hline

			{1,000}        & {303,165}          \\ \hline
			{10,000}      & {3,028,002}     \\ \hline
			
		\end{tabular}
	}
\vspace{-20pt}
\end{table}
Next, we determine the optimal scale $W$ for solving the particle transport problem on STU, balancing accuracy and simulation cost. Fig.~\ref{figMCW} compares the squared error between the solutions for $W=10, 100, 1,000$ and the reference solution for $W=10,000$. Table~\ref{tab4} lists the particle tracking counts for these scenarios. The squared error decreases to 1e-5 for $W=100$ to $W=1,000$, with the latter requiring 10 times more tracking steps. Therefore, $W=100$ is selected as the optimal simulation scale.

The MC problem is solved using STU by matching the particle distribution with the PDF defined by the state transition matrix. The generation of random numbers sequences represents the history of particle tracking.

\begin{figure}[t]
\vspace{-15pt}
\centering
\includegraphics[width=0.98\linewidth]{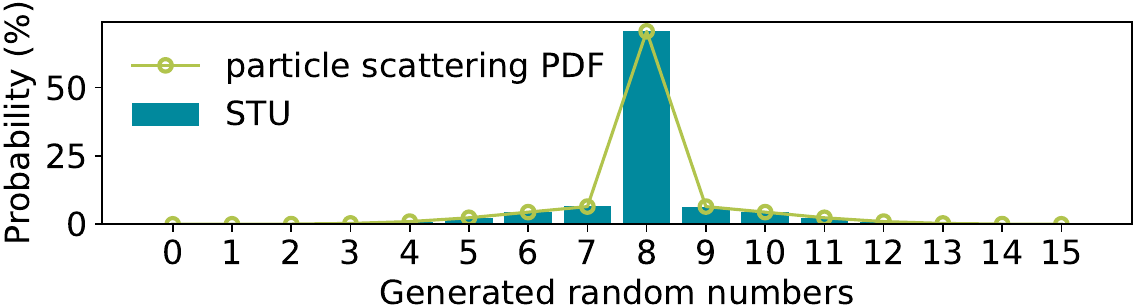}
\caption{Comparison of random events sampled by STU with the PDF of particle scattering direction distribution.}
\label{figRNs}
\end{figure}

Fig.~\ref{figRNs} shows the frequency distribution of 350,000-length random numbers generated by STU (blue bars), alongside the discrete PDF of the scattering direction distribution defined by the Poisson process $P(t)$ and Equation (\ref{equG}) (yellow line). The initial direction of the particles is set to $i=8$ to cover the full distribution range, with other directions calculated by offset. The particle tracking process provides the solution of the angular flux density, shown in Fig.~\ref{fig19}(b). The squared error of the solution, with $W=10,000$ as the approximate solution, is given in Fig.~\ref{fig19}(c). A squared error below 1e-4 suggests that the STU can effectively solve the particle transport problem.

\begin{figure}[t]
\centering
\vspace{-10pt}
\includegraphics[width=0.98\linewidth]{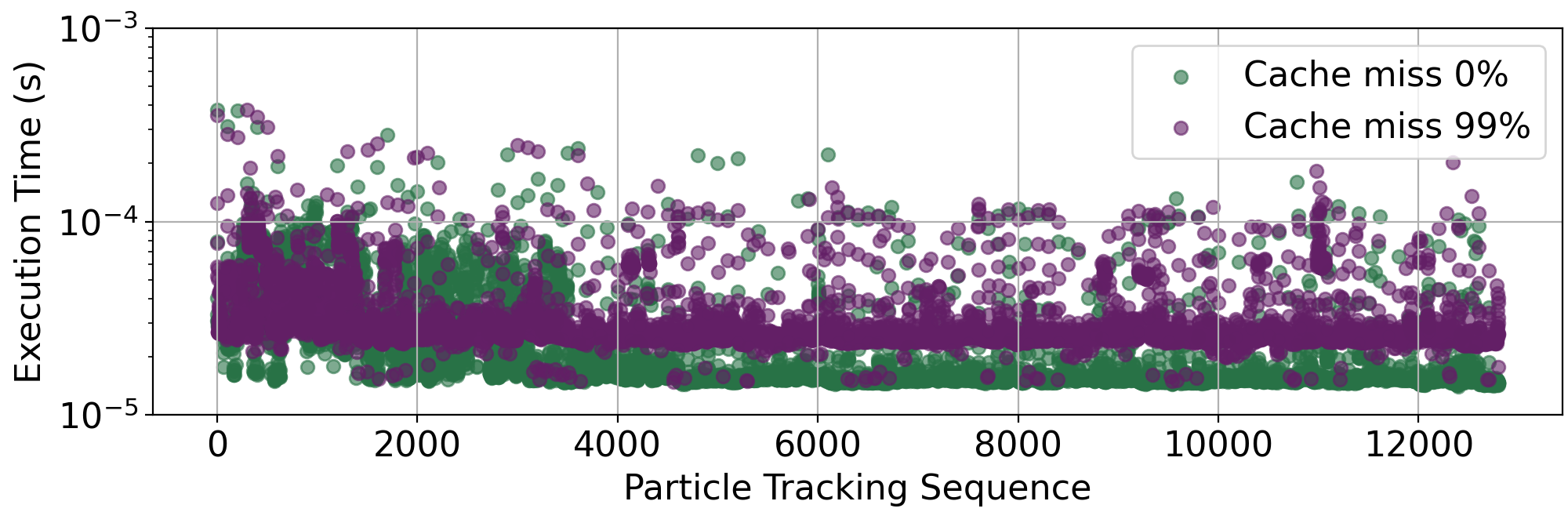}
\caption{The CPU time for a single particle tracking includes both low cache miss rate and high cache miss rate as boundary cases.}
\label{fig28}
\vspace{-15pt}
\end{figure}

To establish a fair comparison with HPC-based MC simulations \cite{matsuoka2023myths}, we model ideal (0\%) and worst-case (99\%) cache miss scenarios. It should be noted that the GPU with 3584 CUDA cores offers only a 55.6$\times$ speedup over the CPU for the transport problem \cite{song2019implementation}, so the CPU serves as the baseline, representing the highest serial performance of a general-purpose processor. Fig.~\ref{fig28} compares CPU execution times per particle for scenarios with 0\% and 99\% cache misses, with realistic tracking times ranging between \SI{21.954}{\micro\second} and \SI{30.390}{\micro\second} per particle. STU achieves a $1000\times$ speedup over a single CPU core tracking by generating random numbers in \SI{21.6}{\nano\second}. This performance advantage, as observed in simulations, suggests that directly generating the target distribution instead of post-processing uniform random numbers may be a promising approach. A potential innovation lies in the ability of the STU to store transition probabilities in a weight selector, which may help reduce cache misses and unify computation with memory access.

In terms of energy efficiency, the STU consumes only \SI{35.1}{\pico\joule} per random sampling, compared to \SI{142.7}{\micro\joule} for CPU-based random sampling, representing up to six orders of magnitude energy improvement over the software baseline as an upper-bound estimate. This aligns with the broader trend that energy efficiency has become a central issue in modern hardware design for scientific computing \cite{hanindhito2026technology1,hanindhito2026technology2}.

\vspace{-10pt}
\section{Related Work}
\label{Related}

In this section, we evaluate NeuroPDE$^+$ against related work. The DTU, which serves as a hardware unit dedicated to the offloading of diffusion processes, is compared to neuromorphic chips \cite{smith2022neuromorphic} that offer similar functionalities. Meanwhile, the STU, a tunable true random number generator designed for sampling from configurable distributions, is evaluated alongside state-of-the-art spin-based true random number generators.

\begin{figure}[t]
\vspace{-15pt}
\centering
\includegraphics[width=0.49\textwidth]{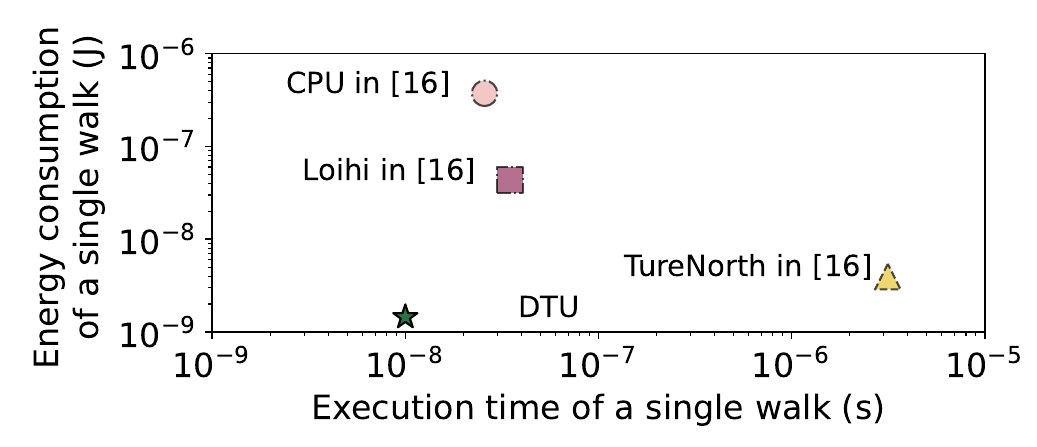}
\caption{Comparison of DTU with a single-core CPU, Loihi and TrueNorth \cite{smith2022neuromorphic} in terms of performance and energy consumption for executing a single MC random walk.
} 
\label{fig14}
\vspace{-15pt}
\end{figure}

Fig.~\ref{fig14} shows the performance of a single random walk on the DTU against other architectures. We measured the execution time and energy of one complete walk, which comprises probabilistic activation, winner-takes-all, and self-inhibition. According to SPICE simulations, the DTU completes each walk in \SI{10}{\nano\second} with an energy consumption of \SI{1.451}{\pico\joule}. For comparison, we use the results of the \textbf{same single random-walk kernel} measured on real CPU, Loihi, and TrueNorth silicon in prior work \cite{smith2022neuromorphic}, which reports actual chip performance rather than simulation estimates. In both cases, the underlying Markov chain model and the random-walk algorithm are functionally identical. Therefore, the comparison in Fig.~\ref{fig14} focuses solely on the kernel latency, host-accelerator initialization and result readout are not part of the kernel and are excluded.

Compared to previous neuromorphic chips, simulation results suggest that the DTU may achieve a speedup of 3.48$\times$ to 315$\times$ in execution time, with a potential energy efficiency improvement of 2.7$\times$ to 29.8$\times$ (reducing energy to 3.35\% to 36.7\% of previous neuromorphic PDE solvers). In summary, CMOS-based neuromorphic architectures such as those in \cite{smith2022neuromorphic} are reported to be more energy efficient than von Neumann architectures, but they may lack performance benefits due to limited inherent randomness. Simulation results from our DTU design suggest that leveraging the physical randomness of emerging spintronic devices could further enhance the performance and energy efficiency of neuromorphic architectures.

\begin{figure}[t]
	\vspace{-10pt}
\centering
\includegraphics[width=0.98\linewidth]{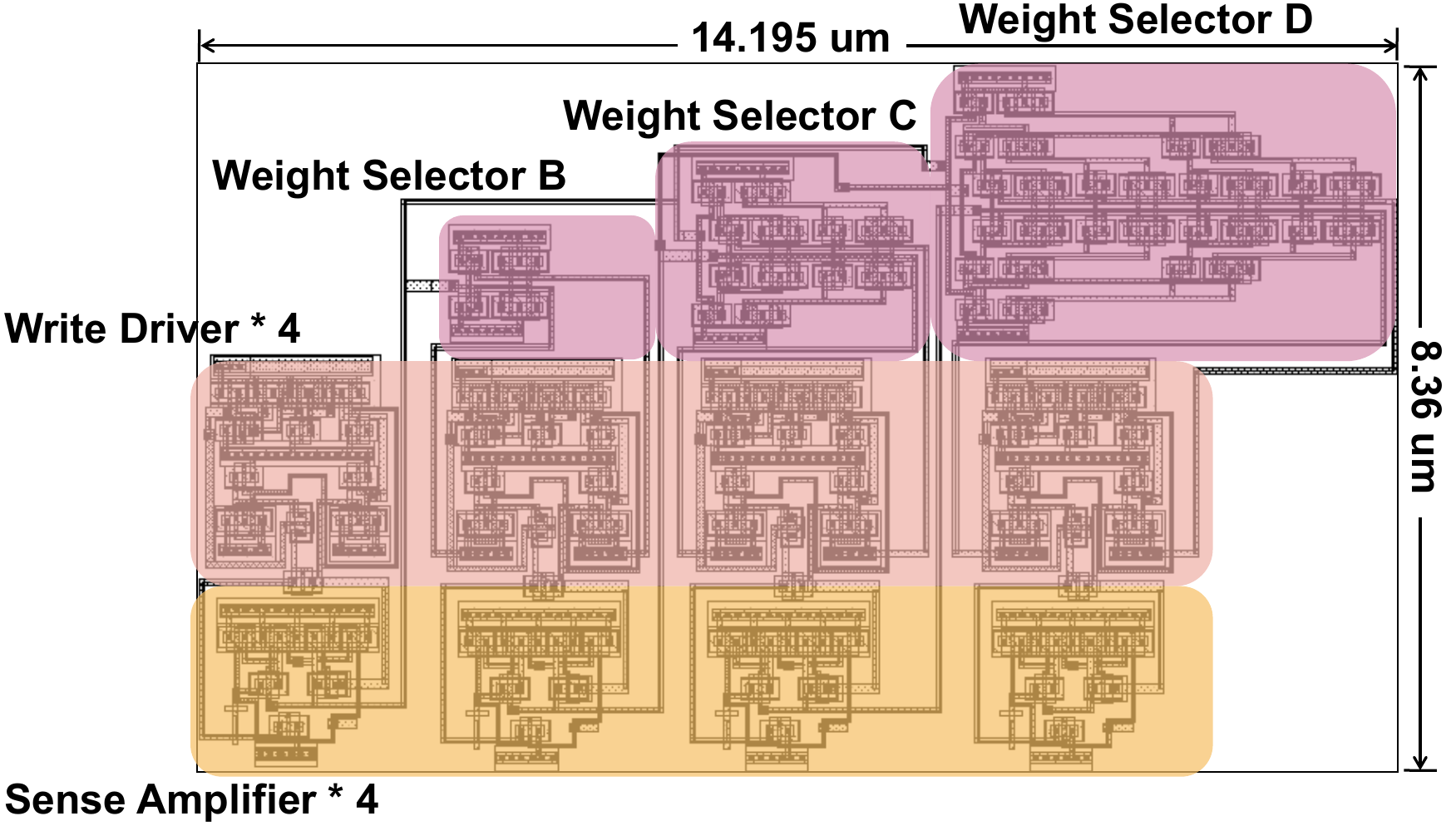}
\caption{STU layout.} 
\label{fig29}
\vspace{-10pt}
\end{figure}

While the STU targets PDE solving rather than random number generation alone, its core sampling engine shares architectural similarity with spin-based TRNGs, making a direct comparison on hardware efficiency and configurability both natural and informative. Table~\ref{tab6} therefore compares the STU with previous spin-based TRNG designs \cite{perach2019asynchronous,amirany2020true,qu2017true,qu2018variation,Fu2023} and recent configurable RNGs \cite{crols2024treegrng,zhang2024probability}. The STU has an estimated random number generation cycle of \SI{21.60}{\nano\second}, with each random number consisting of 4 bits, yielding a bit throughput of approximately \SI{185.19}{\mega\bit}/s. The on-chip area, evaluated using a layout design in the GPDK045 technology (Fig.~\ref{fig29}), suggests that the STU, composed of four MTJ sets, would occupy approximately \SI{118.67}{\micro\metre^2}. Power consumption, derived from transient simulations, is estimated to be \SI{35.05}{\pico\joule} per 4-bit RN, with an average energy consumption of \SI{8.76}{\pico\joule}/bit per individual bit, despite minor differences in the weight selector structures.

Compared to previous spintronic-based TRNG designs \cite{perach2019asynchronous, amirany2020true, qu2017true, qu2018variation, Fu2023}, STU achieves higher throughput than all other works, except RHS-TRNG \cite{Fu2023}. The lower throughput of STU is due to its relaxation of the write time to cover all switching probabilities. The increased power and area consumption are mainly attributed to the conditional probability tree used to control the PDF distribution, as well as technology differences, which make direct comparisons less fair. Overall, STU offers high throughput with comparable area and power consumption compared to previous spintronic TRNGs with fixed output probabilities. Our comparison focuses primarily on RNGs with configurable PDFs.

Although TreeGRNG \cite{crols2024treegrng} offers higher throughput, LFSRs are not true random entropy sources, and their stored probability values suffer from quantization precision limitations due to the digital logic design. In terms of area, TreeGRNG occupies a substantial chip area of \SI{1172.45}{\micro\metre^2} (normalized to \SI{0.895}{\micro\metre^2} per NAND gate \cite{Aguiar2019}), resulting in a per-bit area of \SI{390.8}{\micro\metre^2}, while STU requires only \SI{29.67}{\micro\metre^2}, representing a 13-fold difference. SOT-based designs \cite{zhang2024probability} generate true random numbers that match the target PDF, with a throughput advantage based primarily on process improvements rather than design. In addition, the paper lacks details on the area and power consumption.  In conclusion, STU is the only existing PVT-tolerant TRNG with PDF configurability.

\begin{table}[t]
\centering
\caption{Parameter comparison between STU and related works.}\label{tab6}
\resizebox{0.49\textwidth}{!}{
\begin{tabular}{
>{}c |
>{}c |
>{}c |
>{}c |
>{}c |
>{}c |
>{}c }
\hline	
         \textbf{TRNG}& \textbf{Entropy}&\textbf{Prob.} &\textbf{True}& \textbf{Throughput} & 	\textbf{Power}& \textbf{Area}  \\ \textbf{design}&\textbf{source}&\textbf{config.} &\textbf{random}&&& \\&&\textbf{Y/N}&\textbf{Y/N}& $\mathrm{Mb/s}$& $\mathrm{pJ/bit}$& $\mathrm{um^2/bit}$\\ \hline \hline

          \cite{perach2019asynchronous} &STT-MTJ&&& 7.7-15.1  &5.7-13.4  &50.6-200.6  \\ \cline{1-2}\cline{5-7}
         
          \cite{amirany2020true} &STT-MTJ&&& 50&1.1  &219 \\ \cline{1-2}\cline{5-7}

          \cite{qu2017true}   &STT-MTJ&N&Y& 66.7-177.8     & 0.6-0.8   &3.8-7.6\\ \cline{1-2}\cline{5-7}
          \cite{qu2018variation}  &STT-MTJ&&& 66.7     & 0.8   &3.84\\ \cline{1-2}\cline{5-7}
               
           \cite{Fu2023}&STT-MTJ&&&  303 &  2.65-5.3 &  14.5-24.29   \\  \hline \hline
           
           \cite{crols2024treegrng}&LFSR&&N&  1094 &  8.3e-4 &  390.8   \\ \cline{1-2}\cline{4-7}

           \cite{zhang2024probability}&SOT-MTJ&Y&Y&  $ < 1000 $ &  - &  -   \\ \cline{1-2}\cline{4-7}
           
          STU in NeuroPDE$^+$&STT-MTJ&&Y&  185.19 & 8.76 &  29.67    \\ \hline
\end{tabular}}
\vspace{-15pt}
\end{table}

\vspace{-10pt}
\section{Discussion and Future Work}
\label{Discussion}

Although promising results have been shown, the current design presents several limitations, particularly when extending to real-world, large-scale problems. First, although the DTU effectively mitigates process and voltage variations, its limited resilience to temperature variations would be exacerbated in practical deployment. Second, DTU and STU were independently evaluated, yet many practical systems couple diffusion and scattering processes, and their synergistic potential remains unexplored. Third, the claimed scalability mainly concerns the potential precision scaling of the STU and DTU, while the associated area overhead and precision upper bound remain unexamined under practical accuracy-cost constraints. Multi-device deployment with inter-device communication is also not yet considered. Finally, the current evaluation relies primarily on simulation. Long-term device reliability issues such as FTJ fatigue and MTJ barrier degradation may manifest in deployed systems in ways not captured by small-scale, short-duration simulations.

Future work will be directed toward addressing these limitations. This includes designing a temperature-robust DTU, investigating joint optimization strategies for the DTU and STU to enhance solving efficiency under resource constraints, examining the accuracy-cost trade-offs of precision scaling, investigating communication strategies for multi-device deployment, and bridging the gap between simulation and manufacturing by incorporating more realistic models of device non-idealities, long-term stability, and noise robustness. These steps are essential for practical implementation.

\vspace{-10pt}
\section{Conclusion}
\label{Conclusion}
This paper presents a MC neural PDE accelerator based on spin-ferroelectric hybrid devices, which leverages the inherent stochasticity of spintronic device switching as an entropy source while utilizing the memristive properties of ferroelectric materials to control probabilities, enabling hardware-based tracking of particle random walks for solving PDEs. Simulation results indicate that, by helping mitigate branch penalties and memory access overhead inherent in conventional processors, both the DTU and the STU may achieve improved particle tracking performance. The successful implementation of this spin-ferroelectric hybrid accelerator suggests that neuromorphic circuits exploiting physical stochasticity could demonstrate a potential approach for efficient MC-based PDE solving.

\vspace{-10pt}
\section{Acknowledgment}
During the preparation of this manuscript, the authors used DeepSeek to improve grammar, syntax, and readability in parts of the text, without generating any scientific content.

\nolinenumbers

\bibliographystyle{IEEEtran}

\vspace{-10pt}
\bibliography{ref}



\vskip -2\baselineskip plus -1fil
\begin{IEEEbiography}
	[{\includegraphics[width=1in,height=1.25in,clip,keepaspectratio]{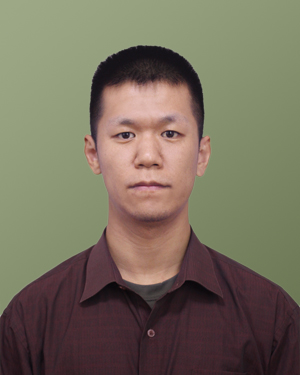}}]{Siqing Fu} received the B.Sc., M.Eng., and Ph.D. degrees in computer science and technology from the National University of Defense Technology (NUDT) in 2018, 2021, and 2026, respectively. His research focuses on domain-specific architecture design and optimization.

\end{IEEEbiography}
\vskip -2\baselineskip plus -1fil

	
	

\begin{IEEEbiography}
	[{\includegraphics[width=1in,height=1.25in,clip,keepaspectratio]{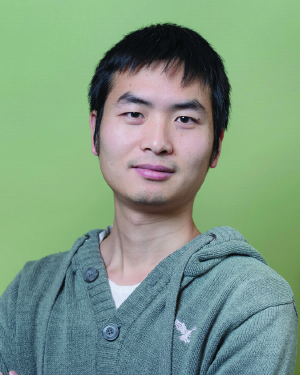}}]{Lizhou Wu}  (Member, IEEE) received the B.Sc. degree from Nanjing University in 2013, the M.Eng. degree from NUDT in 2015, and the Ph.D. degree (Hons.) from Delft University of Technology in 2021. He is currently an assistant professor with NUDT. His research interests include emerging computing paradigms based on non-volatile memory and heterogeneous systems. He received the IEEE TTTC E.J. McCluskey Doctoral Thesis Award in 2021, the Best Paper Award at DATE'20, and three best paper nominations.
\end{IEEEbiography}
\vskip -2\baselineskip plus -1fil

\begin{IEEEbiography}
	[{\includegraphics[width=1in,height=1.25in,clip,keepaspectratio]{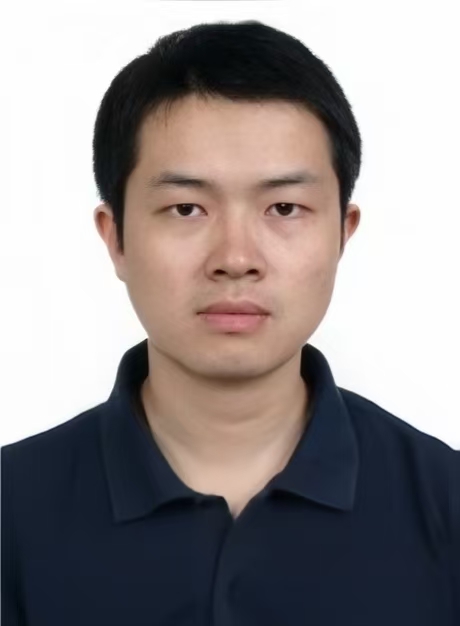}}]{Tiejun Li} received the B.Sc. and Ph.D. degrees in computer science and technology from the NUDT. He is currently a chair professor with the College of Computer Science and Technology, NUDT. He is the Director of a series of research projects.
His research interests include high performance computing and microprocessor architecture. 
	
\end{IEEEbiography}
\vskip -2\baselineskip plus -1fil

\begin{IEEEbiography}
	[{\includegraphics[width=1in,height=1.25in,clip,keepaspectratio]{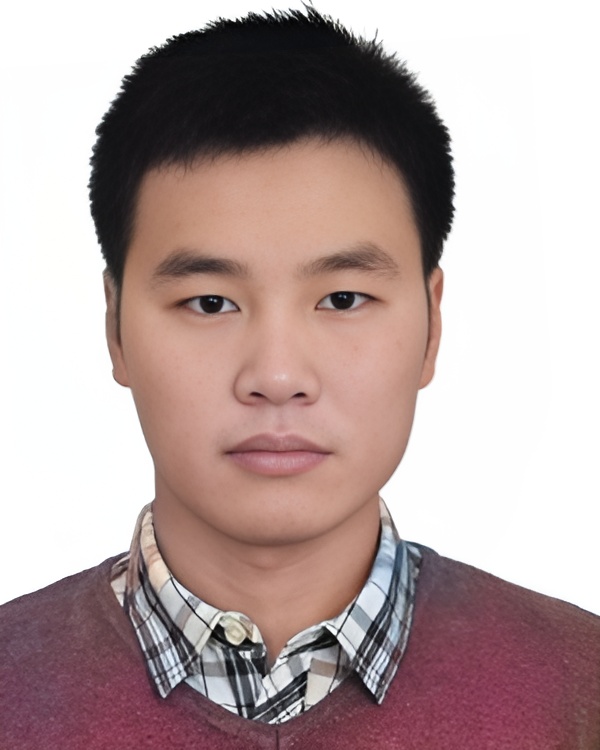}}]{Xuchao Xie} received the PhD degree in computer science from the NUDT, China, in 2018. Currently he is an assistant professor with the College of Computer, NUDT, China. His current research interests include high performance computing, file and storage systems with non-volatile memory, solid-state drives, and shingled magnetic recording drives.
	
\end{IEEEbiography}
\vskip -2\baselineskip plus -1fil

\begin{IEEEbiography}
	[{\includegraphics[width=1in,height=1.25in,clip,keepaspectratio]{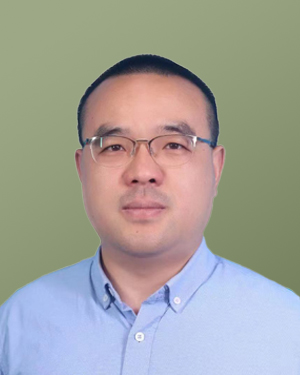}}]{Sheng Ma} received the B.Sc. and Ph.D. degrees in computer science and technology from the NUDT in 2007 and 2012, respectively. He visited the University of Toronto from September 2010 to September 2012 as a co-supervised PhD student. He is currently a professor with  NUDT. From December 2012 to December 2021, he successively served as an assistant professor and an associate professor at  NUDT. His research interests include microprocessor architecture, AI accelerators, and on-chip networks. 

\end{IEEEbiography}

\vskip -2\baselineskip plus -1fil

\begin{IEEEbiography}
	[{\includegraphics[width=1in,height=1.25in,clip,keepaspectratio]{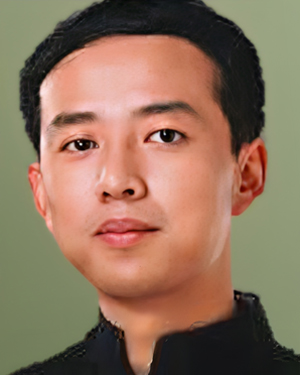}}]{Jianmin Zhang} received the B.Sc., M.Eng., and Ph.D. degrees in computer science from the NUDT, Changsha, Hunan, China, in 2001, 2003, and 2008, respectively. He is currently an associate professor in computer science at NUDT. 
	His major research ﬁelds of interest include high performance computer architecture, and interconnect network.

\end{IEEEbiography}
\vskip -2\baselineskip plus -1fil

\begin{IEEEbiography}
	[{\includegraphics[width=1in,height=1.25in,clip,keepaspectratio]{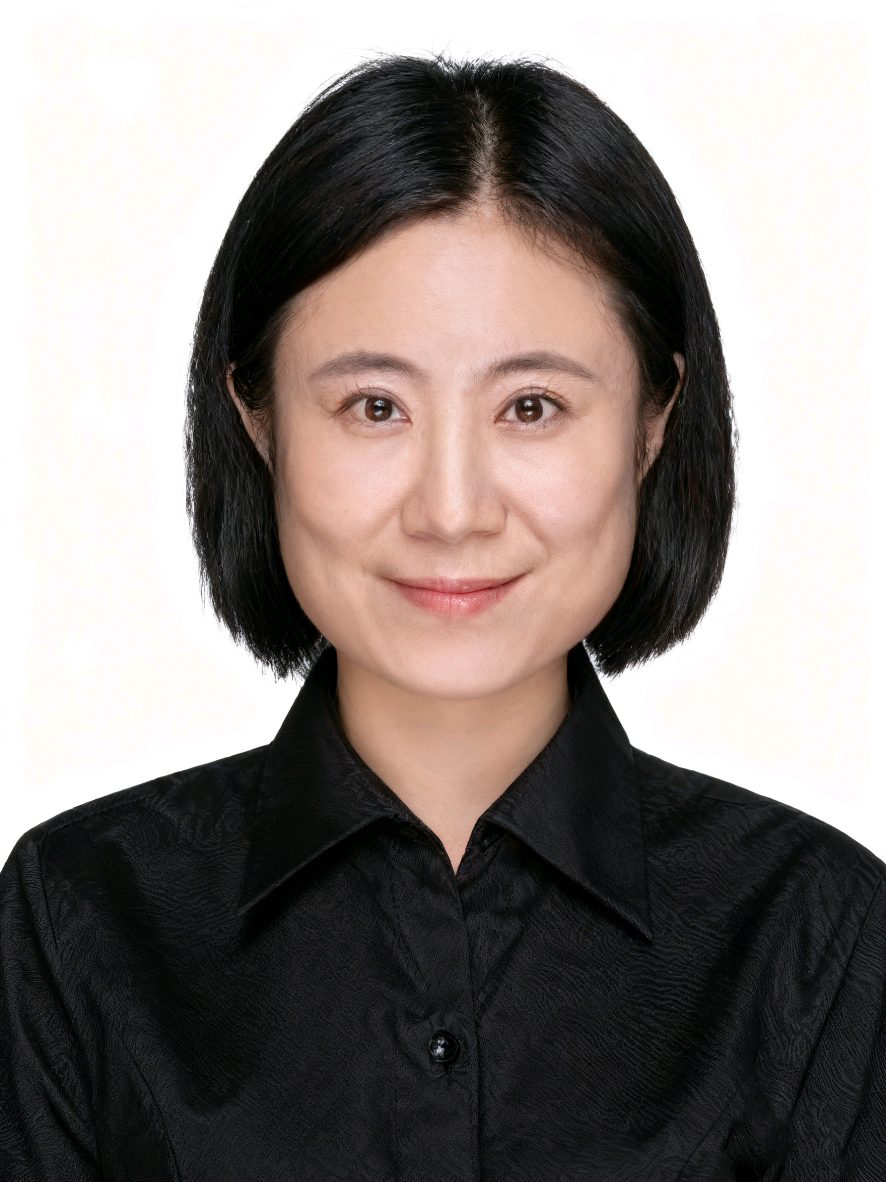}}]{Wei Chen} is currently a Professor with the College of Computer Science and Technology, NUDT. Her current research interests include computer architecture, artificial intelligence, and computer vision.

\end{IEEEbiography}

\vskip -2\baselineskip plus -1fil

\begin{IEEEbiography}
	[{\includegraphics[width=1in,height=1.25in,clip,keepaspectratio]{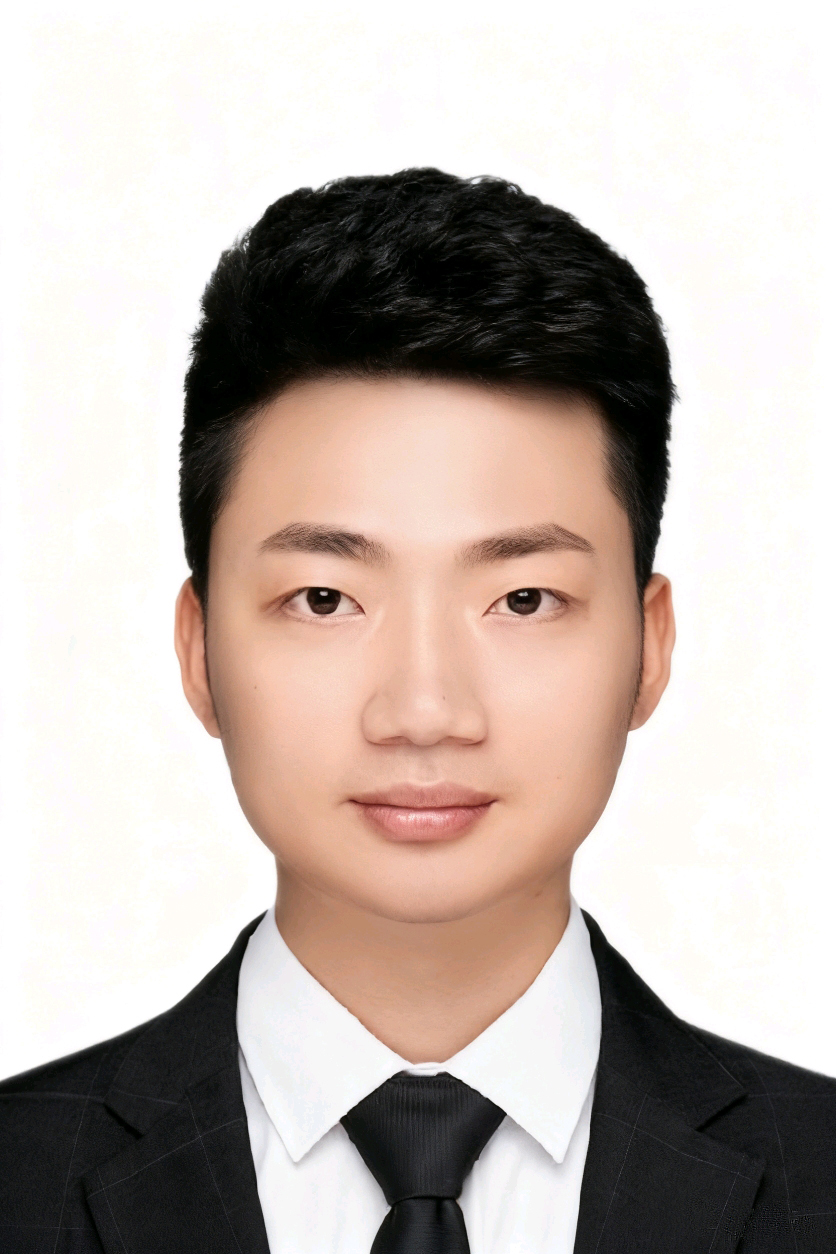}}]{Yunping Zhao} received the B.S. and Ph.D. degrees from the NUDT, in 2020 and 2024, respectively. He is currently a Postdoctoral Researcher with the College of Computer Science and Technology, National University of Defense Technology. His research interests include deep neural networks, processing-in-memory computing, and brain-inspired neuromorphic computing. He has published more than ten papers in TCAD, TACO, TODATES, and ICCD.

\end{IEEEbiography}

\vfill
\end{document}